\documentstyle[12pt,epsf]{article}

\newcommand{\sectie}[1]{{\section{#1}}{\setcounter{equation}{0}}} 

\begin{document} 
\title{Spin order on pyrochlore lattices studied by the Self-Consistent
Cluster Approach.} 
\author{Wiebe Geertsma, \\ 
UFES-CCE/Depto de Fisica,\\ 
Fernando Ferrari s/n, Campus Goiabeiras,\\ 
29060--900 VITORIA--ES (Brasil);\\ 
} 
\maketitle 
 
\begin{abstract}

In this paper we present   a study of  magnetic order  of  quantum
Heisenberg spins  on {\em pyrochlore} type lattices: a lattice  of
corner  sharing   tetrahedra.     We  use   the Cluster  Variation
Approximation   with the  tetrahedron as  basic cluster.  We  find
that  in case of  anti-ferromagnetic   nearest--neighbour exchange
there is no long--range  spin order at any  finite temperature for
any spin ${\bf s}$.  For  ferromagnetic exchange  and $s=1/2$  the
spin system remains in a non-magnetic  phase for all temperatures,
while for $s  > 1/2$   there  is a  finite order temperature.   We
derive   the ground state phase diagram as a function of  exchange
anisotropy and   tetragonal   anisotropy  and find  regions     of
re--entrant  behaviour.  Magnetic  fields  can  induce one or more
magnetic transitions.   Schottky--like   anomalies   appear in the
specific  heat, with on  top signs  of re--entrant  behaviour.  The
limits of this Cluster Variation Approximations are discussed.

\end{abstract}
 
\vfill PACS: 75.10, 75.30E, 75.30H.\newline 
(Submit to Phys. Rev. B )\newline 
Keywords:theory, spin order, magnetism, insulators, .\newline 
Short title: Spin order on Pyrochlore lattices.  
%%%%%%%%%%%%%%%%%%%%%%%%%%%%%%%%%%%%%%%%%%%%%%%%%
%%%%%%%%%%%%%%%%%%%%%%%%%%%%% 
 
\sectie{INTRODUCTION. \label{section:-introduction}}

In this paper   we report results  for  the critical  temperature,
susceptibility,     magnetization  and specific  heat  for Quantum
Heisenberg  spins ($s=1/2$ to $s=7/2$)  on the  highly  frustrated
pyrochlore    lattice using the Cluster  Variation   Approximation
(CVA). This pyrochlore  lattice  is a lattice with  Corner Sharing
Tetrahedra    (CST) which   has   the highest  known   degree   of
frustration according to Lacorre's definition~\cite{Lacorre}.  

In the last  decade  the  properties  of spins  on this pyrochlore
lattice  (see figure~\ref{figure:-pyrochlore-structure})      have
received  much attention because it promises  to be  a realization
of a 3D Quantum  Spin Liquid  (QSL)  with  possibly exotic  ground
states.  Investigated  Quantum ground states are for example based
on singlet valence bonds, leading    to a Resonating Valence  Bond
(RVB) system,    introduced   by  Anderson~\cite{Anderson-RVB-triangles,
Fazekas-Anderson-1974} for a 2D triangular  lattice,  or a Valence
Bond Solid/Crystal,   or  order based on a singlet tetrahedron
(=RVT)  or
plaquette state, or the formation  of   a ground state  with  long
range chiral order.

One of the basic   features  of such frustrated  spin lattices  is
their  extensive   residual or zero--point  entropy  at $T=0$,  or
formulated    in another   way - a macroscopic degenerate   ground
state,  with a degeneracy $\propto a^N$, where $a>0$ and $N$ is of
the   order  of  Avogadro's   number.  The consequences    of this
degeneracy    are many:    1) a large decrease    of the  ordering
temperature  $T_c$, such that $f = \Theta_{CW}/T_c  \gg 1$,  where
$\Theta_{CW}$        is       the  asymptotic         Curie--Weiss
temperature~\cite{review-Ramirez,  review-Schieffer-Ramirez};   in
the region below the classical Mean--Field transition  temperature
the spins enter in  a highly correlated    quantum state, which in
the  case  of classical    spins       is  called  a ``cooperative
paramagnet''~\cite{Review-Geo-Frus-Moessner}, and a possible  spin
freezing     below      $T_F$    into      a    Spin         Glass
phase~\cite{Loc-latt-disorder-frus-SG-Y2Mo2O7,        ZnCr2O4-QAF,
SG-Ferro};  2) a shift of spectral weight  to  low   energies from
high    energies         of  the   order     of      $\Theta_{CW}$
\cite{Geo-frus-Spin-Ice-neg-exp};  3) Schottky--type  anomalies in
the specific    heat    in  case there   are   no   states  in the
singlet--triplet  gap  and   a specific   heat $\propto  T^n$ with
singlet  states   in  the  singlet--triplet      gap,    while the
susceptibility  is always thermally activated  ; 4) small external
perturbations  can lift (part) of the ground state  degeneracy and
thereby  may radically  change the type of (exotic) ground  state:
for example lattice distortions related to  a Jahn--Teller  effect
\cite{Pyr-JT-Spin,order-by-distortion}            bond    disorder
\cite{order-by-disorder-bond-disorder},   field     induced  order
\cite{CF-Mag-Field-Liquid-Gas,         Dy2Ti2O7-GS-Magnetic-Field,
Reentrant-Mag-SL} or the formation of a   spin  glass   phase;  5)
order-by-disorder:  such  small perturbations can also  be quantum
and/or   thermal  fluctuations (entropic ground  states selection)
\cite{order-by-disorder-Villain,  Villain-o-b-d, Harm-SW-QAF-pyr};
6) anomalous,    non-universal,        critical         properties
\cite{Crit-Prop-Pyr-AF};          7)    peculiar  spin    dynamics
\cite{Critical-Dynamics-Frustration};   8) in case  of Ising spins
quantized along the local $<\!111\!>$ axis  a cubic spin--ice type
ground state~\cite{Spin-Ice-Ising,  Ice-Rules-Pyr}, which, in case
that the dipole--dipole   interactions   are stronger    than  the
exchange    interactions,           is        described         as
Dipolar--Spin--Ice~\cite{Dipolar-Spin-Ice-Numerical,
Dipolar-Order-Geo-Frus-AF,                Dipolar-Origin-Spin-Ice,
Dipolar-Origin-Spin-Ice-MFT}

Reviews of experimental work on the susceptibility  of Rare--Earth
Stannate        pyrochlores          can        be     found    in
\cite{Review-Susceptibility-RE2Sn2O7},        for    the  magnetic
properties  of compounds  with   a pyrochlore   spin lattice    in
~\cite{review-experimental-Harris-Zinkin},       and  for the bulk
magnetization     of  Rare--Earth       Titanate       pyrochlores
in~\cite{review-Magnetization-RE2Ti2O7,   review-Ramirez}. Reviews
of  theoretical     work   are   given   by Bramwell and   Gingras
\cite{review-spin-ice},  Moessner~\cite{Review-Geo-Frus-Moessner},
and a review by Lhuillier   and   Misguich~\cite{review-frus-magn}
where  the pyrochlore  system is discussed in  the  perspective of
the general class of frustrated quantum spin systems. 

Anderson~\cite{Anderson-Spinel-1956}    started the study of order
on   the pyrochlore  lattice  and compared  it with the positional
disorder  of  H in hexagonal ice.  He predicted   the absence   of
Ising-type  order   down to $T=0$    in  case of antiferromagnetic
nearest--neighbour   interactions   for  spins quantized  along  a
tetragonal          axis,     calling     such         a    system
``Cubic--Ice''~\cite{Pauling, Baxter-Exact}.    We  can sum up the
present  understanding     of the  Classical     Antiferromagnetic
Heisenberg spin $s=1/2$  on the pyrochlore  lattice  as   follows:
this spin system does not order   from    the  point of  view   of
classical spins:   the highest $J({\bf   q})$ modes are  two--fold
degenerate     without      any   dispersion~\cite{CAF-n-comp-vec,
CHAF-pyr-Reimers, SL-Pyr-CAF},  meaning that  below the mean-field
ordering  temperature $T_{c,MF} = 4/3J_{nn}$ the spin system moves
in these  modes, thereby   using  only 1/2 of the total number  of
degrees      of   freedom.          Moessner                   and
Chalker~\cite{Class-geo-frus-AF-low-T, CHAF-pyr-SL} showed   by an
explicit construction procedure      that   all  ground      state
configurations are connected, without any  energy barrier. However
certain correlations  remain, which  give rise to certain features
in the  neutron   scattering:  a few long-range  correlations   in
high--symmetry directions is build into the ground state. 

Various authors have studied possible  ``order--by--disorder''  in
this lattice, introducing  thermal or  quantum fluctuations on top
of the    classical          ground     state,       see       for
example~\cite{Harm-SW-QAF-pyr}.    Most authors, probably starting
with  Reimers~\cite{CHAF-pyr-Reimers},     first  diagonalize  the
Heisenberg    antiferromagnetic        spin   Hamiltonian     with
nearest--neighbour   exchange on an  isolated tetrahedron and then
construct the  pyrochlore    lattice  with   these non-overlapping
tetrahedra,  and introduce    the interaction    between     these
tetrahedra   as    a  perturbation:  $\lambda   J_{nn}$.       The
tetrahedron--tetrahedron    interaction has been calculated  up to
3rd order    in    $\lambda$.           Garcia--Adeva          and
Huber~\cite{MFQ-Tetr-Pyr}     used the  most simple  approach   by
neglecting   these tetrahedron--tetrahedron interactions: in  this
way  long--range  order is excluded   from   the   outset.     The
asymptotic ``Curie--Weiss''  temperature found is 1/2 of the value
obtained from  the canonical mean field approximation.  Canals and
Lacroix \cite{QSL-HAF-Pyr,QAF-3D-Pyr} performed  a density  matrix
expansion    up to $\lambda^3$  and found that  at this  order the
degeneracy is weakly lifted ($\propto 10^{-6}J$) giving rise  to a
collinear AF spin  structure:   the  spins along  $<\!110\!>$  are
parallel    while  nearest--neighbour   perpendicular  $<\!110\!>$
crossings are antiparallel.  The correlation length is  at most of
the order of one  lattice   spacing.     Their   approach  can not
elucidate the character   of the ground state: whether it has long
range order in one of the exotic ground states mentioned above. 

Several  authors    have discussed this possibility by calculating
the tetrahedron--tetrahedron  interactions in the restricted space
of  the two--fold degenerate singlet states and 3--fold degenerate
triplet states  of the tetrahedron.  The  latter are essential for
inter--tetrahedron  interactions.    Although    starting     from
different points  of view, the ground  state  of this  spin system
seems    to be some singlet--dimer covering  of   the pyrochlore
lattice: Tsunetsugu~\cite{AFQ-pyr} used a chiral ansatz  and found
vanishing chirality: the ground state is  a four sublattice system
of tetrahedra with  singlet  dimer order on three sublattices each
occupied by one of the possible  tetrahedron singlet dimer states,
and the 4th sublattice is locally disordered.The   ground state of
the  singlet dimer lattice can be  characterized as a Valence Bond
Solid.   Singlet excitations  will  appear in the singlet--triplet
gap:    these  are oscillations   in  the dimer pattern. A similar
ground    state   has   been   found   by    Eljahal    {\it    et
al}~\cite{Tetrahedra-QHAF},                  by     Isoda      and
Mori~\cite{Pyrochlore-Plaquette-Isoda-Mori}.     A  dimer covering
was  already        discussed           by   Harris     {\it    et
al}~\cite{Harris-Berlinsky-Bruder-1991}.   Recently Berg  {\it  et
al}~\cite{Pyrochlore-Singlet-Exc}  pushed  this  approach one step
further and used the two--fold degenerate  singlet ground state of
a super tetrahedron  - consisting  of 4 tetrahedra  connected each
by different  bonds - as a basis for  a perturbation  calculation.
To each  super tetrahedron is assigned  a pseudo spin, these super
tetrahedra  form   a  bcc lattice.  From   a Monte--Carlo   study,
treating these pseudo spins as classical entities.  They find that
the ground   state  is  ordered  anti ferromagnetically,      with
ferromagnetic layers.

Adeva-Garcia                                        and
Huber\cite{Classical-generalized-constant-coupling}   have studied
the magnetic properties  of spins on the pyrochlore  lattice using
the constant--coupling  approximation  and recently  applied it to
ZnCr$_2$O$_4$\cite{ZnCr2O4-QAF}.     The  same authors studied the
critical   properties  of these pyrochlore  spin  systems using  a
Renormalization                                   Group
appraoch\cite{Effective-Renormalization-Group}.

When one looks into the pyrochlore structure   one finds that each
atom on  this   lattice shares six planar hexagons,  two   in each
$<\!111\!>$  direction,   of which   each of the  three  pairs  of
hexagons are  edge sharing.  The hexagons in a $(\!111\!)$ plane
form a Kagom{\'e} lattice.  So this pyrochlore   lattice  also may
be characterized  as  a lattice of  inter-penetrating   Kagom{\'e}
lattices.  

Let  us   now  turn  our attention     to {\it   nn} ferromagnetic
interactions.  There is  at first sight no reason why the spins on
a lattice of CST would not order  at finite  temperature.    First
consider the  case that the spins are Ising spins pointing   along
the local $<\!111\!>$.     In  that case there are six  degenerate
states  with  a two-spin-in  two-spin-out   configuration  on each
tetrahedron~\cite{Spin-Ice-Ising,Ice-Rules-Pyr}      and  the spin
states can   be mapped  on a model describing   the   two possible
positions  of Hydrogen in the ice structure  \cite{Pauling}.  This
leads to so--called   ice rules for   the possible low temperature
spin configurations.    A clear exposition of these  rules and its
consequences    can be  found        in  \cite{    Spin-Ice-Ising,
Ice-Rules-Pyr,Spin-Ice-Ho2Ti2O7}.   Because of this  ground  state
degeneracy,  the entropy at $T=0$ K is finite, and the macroscopic
ground state of this spin system is called spin--ice.

In the present paper we apply the Cluster Variation  Approximation
(CVA) for  the configurational  entropy,  in which one divides the
lattice       in        a         set       of         overlapping
clusters\cite{Ziman-CVA,CVA-Ising} (basis cluster)  to study   the
magnetic properties of Quantum Heisenberg    spins on a pyrochlore
lattice. We use  the tetrahedron  as basis  cluster.  We   include
tetrahedral   anisotropy,   and an anisotropy  in  the    exchange
interactions  caused  by a tetragonal  distortion of the  lattice.
Single--ion  anisotropy and  dipole--dipole     interactions   are
discussed briefly. 

Recently  this approach has been   applied   to the    problem  of
classical Ising spins  on a pyrochlore  lattice by  Yoshida  et al
\cite{CVM-spin-ice}.    The properties of  this  classical   Ising
model are quite different from the ones we report  in  this paper,
especially the ground state degeneracy.

The CVA is simply the Mean Field approximation  in  case the basic
cluster is a single  site.  In  case the whole crystal is taken as
the basic  cluster  the approximation becomes exact.    In between
one has  to make an  intelligent choice for the  basic cluster: we
have   chosen  the tetrahedron    as   the  basic    cluster.  The
tetrahedron  is the smallest unit possessing all the  symmetry  of
the lattice,  it  is the unit which seems to  contain    the basic
magnetic properties    of  the spin  system: the high  frustration
against order  in case of AF order.  Another appealing property is
that one  can easily diagonalize the spin Hamiltonian  for such  a
cluster,   and the  Clebsch--Gordan  coefficients are also  easily
calculated.  

The hexagons    next    to    the triangles,     are  the smallest
self-avoiding   closed loops on  the pyrochlore   lattice.  So any
description   based only on the  tetrahedral    clusters, means we
neglect   these   hexagon   closed   loops.    In this sense   our
approximation       is  similar      to  that     for   a   Husimi
tree~\cite{Husimi-SL}   lattice,  although  our approximation does
not suffer  from the rather  peculiar increase  in  the  number of
surface clusters, see also~\cite{Cayley-tree-FI-Glass}.

This      paper     is  organized     as  follows.    In   section
\ref{section:-theory} we introduce  and discuss qualitatively    a
Hamiltonian describing   the spins on this pyrochlore lattice.  We
transform the site Hamiltonian   to a   Hamiltonian for tetrahedra
(\ref{subsection:-Tetrahedron-Cluster}).            In     section
\ref{subsection:-5tetrahedra}    we discuss the spectrum of  a  16
site     cluster.          We       then    derive    in   section
\ref{subsection:-independent-tetrahedra} the susceptibility  for a
pyrochlore lattice covered with non interacting tetrahedrons.   In
section \ref{subsection:-CVA-Theory}  we give some   basic details
of how we apply the CVA to this  Heisenberg Quantum  system.    We
derive  an    expression    for  the    (linear)    susceptibility
(\ref{subsection:-susceptibility}).     In    the    next  section
\ref{section:-results-discussion}    we present first the critical
temperature  (\ref{subsection:-Tc})     and  we present   the zero
temperature  phase diagram   for  ferromagnetic  and  AF {\it  nn}
interactions (\ref{subsection:-GS_s=1_2}), and then present    the
principal numerical  results on the susceptibility,  magnetization
as a functioon     of  magnetic     field      and     temperature
(\ref{subsection:-M-T-H-Ferro}).           In   a final    section
\ref{section:-summary-conclusions}   we give a brief  summary  and
conclusions.  

\sectie{THEORY.\label{section:-theory}}  

The general  Hamiltonian  to describe spin--spin  interactions  is
the  Heisenberg    Hamiltonian.    We consider  nearest--neighbour
($J_{nn}$), and briefly we  will also discuss   qualitatively  the
consequences   of next--nearest--neighbour      ($J_{nnn}$),   and
next--next--nearest--neighbour          ($J_{nnnn}$)      exchange
interactions.   We assume that each of these interactions does not
depend on the details  of  the interaction path  in the pyrochlore
structure: {\it nn}, {\it nnn},  and  {\it nnnn} indicates  by how
many cation--cation links these  cations  are separated.  From  an
analysis of ESR spectra  of Cr$^{3+}$--doped  spinel ZnGa$_2$0$_4$
Henning~\cite{Henning} derives  the Cr$^{3+}$--Cr$^{3+}$  exchange
constants on this spinel  B sublattice.   These exchange constants
show only small  deviations -- of about  10 \% -- from the average
value  of each type  of  exchange  path.  We also     consider the
possibility  of local  single--ion    trigonal     and  tetragonal
anisotropy.  In  case of the single--ion    anisotropy the   local
trigonal field  is along a local $<\!111\!>$ axis pointing to the center
of the tetrahedron.    The  tetragonal  anisotropy can be due to a
tetragonal  distortion, due to uni-axial pressure,   of  the cubic
pyrochlore or  spinel structures.  It  will be  accompanied  by  a
change in the    two   {\it  nn}   exchange  interaction    in the
$\{\!100\!\}$    planes with  respect to  the other four  {\it nn}
exchange interactions in the $\{\!111\!\}$  planes    (see  figure
\ref{figure:-tetrahedron}).  The Hamiltonian for this  spin system
in a magnetic field ${\bf H}$ reads: 
\begin{eqnarray}   
H&  = &J_{nn}\sum_{i,j}   {\bf s}_i  \cdot  {\bf s}_j   +
J_{nnn}\sum_{i,j} {\bf s}_i  \cdot  {\bf s}_j + J_{nnnn}\sum_{i,j}  {\bf s}_i    \cdot
{\bf s}_j   + \Delta\sum_{i,j;pairs}{\bf s}_i    \cdot   {\bf s}_j  \nonumber \\  && +
D_{0s}\sum_i   ({\bf s}^{\prime\;   2}_i       -   3s^{\prime\;   2}_{iz})     +
D_T\sum_{\alpha}({\bf S}_{\alpha}^2   - 3S_{\alpha\;   z}^2) + \sum_i g\mu_B {\bf
H}\cdot{\bf s}_i,   
\label{equation: basic   Hamiltonian}     
\end{eqnarray}
The  fourth contribution  is  a sum over the exchange interactions
in the $\{\!100\!\}$ planes.  We reserve   capitals ${\bf S}_{\alpha}$
for the size of the cluster  (tetrahedron) spins and small  ${\bf s}$
for the local atomic spins. The  magnetic  field  ${\bf H}$ in the
last term    is    along   a   tetragonal      axis (see    figure
\ref{figure:-tetrahedron}). 

The $ {\bf s}^{\prime}$ in the single--ion  contribution     $D_{0s}$
term indicates  quantization of the spin along  the local trigonal
$<\!111\!>$ axis.  In case the single--ion  anisotropy  $D_{0s}>0$
the local   $<\!111\!>$   axis   is  the easy axis,  while in case
$D_{0s}<0$,  the spin  rotates in the $x^{\prime}y^{\prime}$ plane
perpendicular   to this axis.  In case $D_{0s}>0$  and much larger
than the exchange  interactions it is often  argued   that one may
approximate       the     spins     as    classical        (Ising)
spins~\cite{Single-Ion-Anis-Pyr}.       In  case  of a  tetragonal
distortion along one of the $S_4$ axes of the tetrahedron one finds
\begin{eqnarray} 
H_{si}&=&\frac{1}{2}D_{0s}(3\cos^2\theta_4-1)
\sum_{i}({\bf s}_{i}^2- 3 s_{iz}^2)\\
&&-3\sqrt{2}D_{0s}\cos\theta_4\sin\theta_4
\left[s_{1z}s_{1+} + s_{1+}s_{1z}- s_{2z}s_{2+} - s_{2+}s_{2z}
+ i(s_{3z}s_{3-} + s_{3-}s_{3z}- s_{4z}s_{4-} - s_{4-}s_{4z})
\right]
 \nonumber \\
&&-\frac{3}{2}D_{0s}\sin^2\theta_4
\left[(s_{1+}^2 - s_{1-}^2) +  (s_{2+}^2 - s_{2-}^2) -(s_{3+}^2 -
s_{3-}^2)- (s_{4+}^2 - s_{4-}^2) \right]\nonumber
\label{equation: single ion  anisotropy}
\end{eqnarray}
where     $\theta_4$           is      indicated      in    figure
\ref{figure:-tetrahedron}.    The single--ion anisotropy  vanishes
in case $s=1/2$.

Next   let us briefly discuss the tetragonal   anisotropy   $D_T$.
When the tetrahedral anisotropy $D_T>0$,  the tetrahedron spin $S$
is aligned along  a tetrahedral  axis,  and in  case $D_T<0$, this
spin   is  rotating  in the plane  perpendicular to  a tetrahedral
axis.  This term can also be written in terms  of the single--site
spins: 
\begin{eqnarray} 
&&(J_{nn}  + D_T)\sum_{<i, j>} s_{i}s_{j}  -
3  D_T\sum_{<i,       j>}s_{iz}s_{jz}    +  2   D_T\sum_i(s_i        ^2  -  3
s_{iz}^2)=\nonumber\\&&  (J_{nn} - 2 D_T)\sum_{<i, j>} s_{iz}s_{jz} + (J_{nn}
+ D_T)\sum_{<i, j>} \left(s_{ix}s_{jx}+s_{iy}s_{jy}\right)  + 2 D_T\sum_i(s_i
^2 -  3 s_{iz}^2).\nonumber \\
&&  
\label{equation:  tetrahedron    anisotropy}    
\end{eqnarray} 
The tetrahedral   anisotropy interpolates    between a Ising model
($J_{nn}=-D_T$) and an $XXZ$ model  to a Heisenberg ($D_T=0$)  and
$XX$ model ($J_{nn}=2D_T$), by  changing the ratio $D_T/J_{nn}$.
In the next   section we  will get more insight   in  this term by
writing the whole Hamiltonian in terms of tetrahedron spins.

%############################################################################
\subsection{The Tetrahedron Approach.\label{subsection:-Tetrahedron-Cluster}}
\subsubsection{The
Hamiltonian.\label{subsubsection:-TheHamiltonian}}   For the   CVA
with  the tetrahedron  as  a basic cluster we need the spin states
of this tetrahedron.   We  define    for tetrahedron $\alpha$  the
total spin as ${\bf  S}_{\alpha}  = {\bf  s}_1 +  {\bf s}_2 + {\bf
s}_3 +{\bf  s}_4$.  For the  $\Delta$ term we introduce ${\bf L} =
{\bf s}_1  + {\bf s}_2$ and ${\bf R} = {\bf s}_3 +{\bf s}_4$.  One
can now  rewrite the exchange part  and   magnetic   field  of the
original Hamiltonian in terms of these tetrahedron spins 
\begin{eqnarray}              
H           &=&
(J_{nn}-2J_{nnn}+3J_{nnnn})\sum_{\alpha}{\bf S}_{\alpha}^2                      +
(J_{nnn} - 
2J_{nnnn})\sum_{(\alpha,\beta)_{nn}}{\bf S}_{\alpha}\cdot{\bf S}_{\beta}   +
J_{nnnn}\sum_{(\alpha,\beta)_{nnn}}{\bf S}_{\alpha}\cdot{\bf S}_{\beta}  
\nonumber \\
&+&\frac{1}{2}\sum_{\alpha}g\mu_B    {\bf H}.{\bf S}_{\alpha}  + 
\Delta \sum_{\alpha}
\left({{\bf L}}_{\alpha}^2  +
{{\bf R}}_{\alpha}^2\right) +
D_T \sum_{\alpha} ( {\bf S}_{\alpha}^2 - 3S_{\alpha z}^2 ) + H_{si} 
\nonumber\\
&&
\label{equation: tetrahedron Hamiltonian}
\end{eqnarray}

The sum is over all tetrahedra in the lattice, including  the ones
which  share a  corner: so in case of the pyrochlore lattice there
are $N/2$ CST.  The sum  over $\alpha$  and   $\beta$ is over  all
tetrahedra under   the condition that  they  are {\it  nn} or {\it
nnn} tetrahedra.   We have dropped terms of type $J\sum_i  s_i^2$,
as these  only add a  constant energy to  the  Hamiltonian.    The
first term  gives the  energy  levels  of an isolated tetrahedron,
with  an  effective      exchange     in     one  tetrahedron   of
$J_{nn}-2J_{nnn}+3J_{nnnn}$.  This correction on $J_{nn}$   should
only be taken   into account when also the the corresponding  {\it
nn} and {\it  nnn} are taken  into account: the   second and third
terms give these interactions between the total spins on {\it  nn}
and {\it nnn} tetrahedra.

We note that several  of these contributions ($D_T$, $\Delta$) also appear in
a summation   of the  dipole interaction   within  a tetrahedron:
\begin{equation}
H_{D} = \frac{D_D}{8}\left[{\bf S}^2 - 3S_z^2\right]+\frac{2
D_D}{3}\left[L^2 +R^2 - 3(L_z^2 + R_z^2)\right] + \cdots
\label{eq:-dipole-dipole}
\end{equation}
where $D_D =  g^2\mu_B^2/R^3$, $R$ is the interatomic distance  in
a tetrahedron.        The  dipole  interaction, diagonal  in   the
tetrahedron  cluster  states,  is  taken  into  account  (except a
contribution  of the form $L_z^2 + R_z^2$), the last  term in eq.
\ref{eq:-dipole-dipole}.   All single  and  double spin flip terms
appearing  in the dipole--dipole interaction  are  not  taken into
account. These contributions  cause the dipole--dipole interaction
neither to commute with $M_s$, nor with $S^2$.

\subsection{The eigenstates of five coupled
tetrahedra.\label{subsection:-5tetrahedra}}

In  order  to discuss  the ground state of a   lattice of CST,  we
calculated the ground  state of a finite cluster  consisting  of 4
outer tetrahedra  each sharing a corner    with  the  same central
tetrahedron: 5T super-tetrahedron.  First let us briefly  consider
the  energy   levels   of the tetrahedron.  These   are  in figure
\ref{figure:-energy-levels-s=1/2}                              and
table~\ref{table:-energy-levels-s=1/2}.   In case of nn AF xchange
interactions the  ground   state is a degenerate  pair of singlets
and in case  of F interactions a quintet state $S=2$. This pair of
singlet states is split by  the $L^2 + R^2$ perturbation  into two
singelts, one with $L=0$,  $R=0$: the two dimer singlet state, and
the other   $L=1$, $R=1$:  the  plaquette  or  tetrahedron Singlet
state. Also the dipole--dipole interaction splits this singlet.

We calculate the ground state energy in case  of  AF interactions.
We proceed  as follows: the Heisenberg Hamiltonian is rewritten in
terms  of the tetrahedron spins $S_i$, for the outer 4 tetrahedra.
The eigenstates   can  easily   be found using  the Clebsch-Gordan
coupling coefficients.     These states are  represented   by  the
following            set     of       quantum             numbers:
$|Q,M_Q,D_1,D_2,S_1,S_2,S_3,S_4,L_1,R_1,L_2,R_2,L_3,R_3,L_4,R_4\rangle$,
where $Q$ and  $M_Q$ are the total spin and magnetic moment of the
outer 4  tetrahedra,  $D_1$ is  the spin  of  the tetrahedra  with
$S_1$ and $S_2$,   and $D_2$  of $S_3$ and  $S_4$.  The $L_i$  and
$R_i$  are defined  above.   This  basis  is diagonal      in  the
Hamiltonian  for  the  exchange  interactions   of  the outer four
tetrahedra.     The   6  exchange interactions     in  the central
tetrahedron have  now to be considered separately.  For $M_Q = 0$,
and  $Q  = 8$, 7, 6,  5,  this can  be done easily.  For the other
$M_Q$ values we have  chosen the  following scheme  in  order   to
determine  the ground state and first excited states:  we choose a
maximum value  for $S_i$,      $D_i$ and $Q$,  for $M_Q  = 0$  and
restrict   the sum over all tetrahedron spins  $S_i$.  That  is, we
restrict the number of singlet--triplet--quintet   excitations  in
the cluster.   Our  results  for the ground state energy  $E_{GS}$
and  singlet--triplet         gap  $\Delta_{ST}$   are  in   table
\ref{table:-energy-levels-quintet-cluster}.       We find that the
ground state is a set  of 16 singlets.  There  is  only   coupling
between       the  four  outer tetrahedra     by  singlet--triplet
excitations.    We    find $E_{GS}     =  -12.82J$     for  our T5
super-tetrahedron,  which is $-0.81J$/site.  This value is smaller
then  obtained for the 16 site   super-tetrahedron   of  Berg {\it
etal}~\cite{Pyrochlore-Singlet-Exc}.  The singlet--triplet gap  is
about $\Delta_{ST} = 1.56J$,  which means  a decrease of about  25
\%  with    respect      to  independent   tetrahedra.          In
table~\ref{table:-groundstate-STgap}  we have collected  some data
from recent calculations of these quantities.

The ground state  degeneracy of the   T5  cluster is determined by
the 2--fold degenerate  ground-state of singlets  of each of the 4
outer tetrahedra.   The  state   on the  central  tetrahedron   is
completely determined  by the   state  on each of the four   outer
tetrahedra.   When   we extend  this  description  to the complete
pyrochlore  lattice --  which  is from  the  point  of view of the
tetrahedra  a  diamond   lattice    -- the  ground    state   is
$2^{N/4}$--fold degenerate for a lattice  of  $N$ sites in case of
AF interactions. 

In our calculations    on the  T5 cluster  we observed  a  partial
lifting of  the degeneracy for certain choices of  the restriction
on the maximum  number of excitations (the size of Hilbert space):
to get the correct degeneracy  one has to diagonalize the complete
matrix.    In this T5 cluster   we    do not find lifting  of  the
degeneracy          due              to          inter-tetrahedron
interactions~\cite{Critical-Dynamics-Frustration}.
Diagonalization    of  only  part of Hilbert space gives incorrect
degeneracies and spurious eigenvalues.

Such singlet  gap-states seem  to  appear  as   a function  of the
cluster size, or in a real system  as a function of the spin--spin
correlation length.   This implies that the extent of spin singlet
states   depends on temperature,        and the   singlet--singlet
excitation spectrum  collapses to a single highly degenerate level
or a narrow band ($\ll J$)   of singlet  states: that is, transfer
of spectral  weight from high to low  frequencies with  decreasing
temperature.  The appropriate correlation lenght  if this  quantum
order increases with decreasing temperature.   At high temperature
($T\ge J$) this correlation    lenght   is  of  the order   of one
interatomic distance.

\subsection{The paramagnetic susceptibility of independent tetrahedra.
\label{subsection:-independent-tetrahedra}}

With  {\it nn} exchange interactions  $J$ only, and  neglecting  the
restrictions due to  corner sharing of spins, one derives easily the  following
expression for the susceptibility\cite{MFQ-Tetr-Pyr}:
\begin{equation}
\chi = \frac{g^2 \mu_B^2}{k_B
T}J \frac{\sum_{S}g_{S}S(S+1)(2S+1)\exp(-E_S/k_B T)}{\sum_S g_S
(2S+1)\exp(-E_S/k_B T},
\label{equation: paramagnetic susceptibility}
\end{equation}
where  $g_S$   is the degeneracy  of a tetrahedral $S$-state  excluding   the
$(2S+1)$  degeneracy which is included explicitly in this   expression.   The
energy of each $S$ state is given by
\begin{equation}
E_S = J S(S+1).
\end{equation}
In the high temperature limit one derives:
\begin{equation}
\chi_{HT} = \frac{g^2 \mu_B^2 }{3k_B T}\frac{G_1}{1-  (G_1 - G_2)J/k_B T},
\end{equation}
where 
\begin{eqnarray}
G_1 & = & \frac{\sum_S g_S(2S+1)(S+1)S}{\sum_S g_S (2S+1)}, \nonumber\\
G_2 & = & \frac{\sum_S g_S(2S+1)(S+1)^2 S^2}{\sum_S g_S(2S+1)(S+1)S}.
\end{eqnarray}
One can now easily calculate the asymptotic Curie temperature:
\begin{equation}
\frac{k_B \theta_{CW}}{J} = \frac{1}{G_1 - G_2}.
\end{equation}
This asymptotic Curie  temperature   is not related to  any long range order.
For the case of $s=3/2$ -- for example Cr$^{3+}$ with d$^3$  -- one finds  an
asymptotic Curie temperature given by:
\begin{equation}
k_B \theta_{CW} = -\frac{15}{2} J,
\end{equation}
which  is  $1/2$ the  value one  would  obtain  from the    usual Mean  Field
approximation!  One finds  the same factor $1/2$ for   the Kagom{\'e} lattice
in case one treats   the lattice as consisting of independent corner--sharing
triangles. 
%############################################################################
\subsection{The Cluster Variation Approximation with tetrahedron 
  as basis cluster.\label{subsection:-CVA-Theory}}

In this section  we give some details  of our calculation  of  the
probability  functions   for   quantum spins for the  case  of   a
tetrahedron  cluster with  $s=1/2$.    The calculation  for higher
spin values is only slightly different.  The CVA  is often applied
to  the  classical  problems of order--disorder    transitions  in
alloys, and the  problem of Ising spins. In the  present   case we
apply  it  to the  problem   of  Quantum  Heisenberg    spins. The
approximation we make  is that there is no coherence   between the
quantum states  on different tetrahedra.  At high temperature this
leads to the  classical Mean Field limit. However we will see that
in  certain  cases below  a certain   temperature $T_{S}$  the CVA
breaks down:  the entropy in this approximation   becomes negative
for $ T < T_S$.

The states  on the tetrahedron are written  as linear combinations
of the determinants: $|m_1 m_2 m_3 m_4|$, where $m_i$  is the spin
moment on $i$. The spin eigenfunctions can be written as:
\begin{equation}
|\Psi(S,M,L,R)\rangle=|\Psi(S,M,\alpha)\rangle =
\sum_{m_1,m_2,m_3,m_4} \langle
m_1,m_2,m_3,m_4|S,M,\alpha\rangle|m_1,m_2,m_3,m_4\rangle,
\label{equation:  Clebsch-Gordan  expansion} 
\end{equation}  
where  $\langle
m_1,m_2,m_3,m_4|S,M,\alpha\rangle$       are   the Clebsch--Gordan
coefficients.  The $\alpha$'s  indicate  one of the eigenfunctions
of the possible degenerate  set  of eigenstates  with spin $S$ and
momentum $M$ with different pairs $(L,R)$.  The restrictions   are
$ |L  - R|  \le  S  \le L + R$, $M =  M_R +  M_L$, $ 0 \le  L  \le
s_1+s_2$, $0 \le R  \le s_3+s_4$, $M_L = m_1  + m_2$, $M_R = m_3 +
m_4$.

The probability that a certain site has magnetic spin momentum $m$ is:
\begin{equation}
x_m = \sum_{S,M,\alpha} a_{m;S,M,\alpha}y_{S,M,\alpha},
\end{equation}
where $y_{S,M,\alpha}$ is the probability that the tetrahedron  is
in one    its  eigenstates    $|S,M,\alpha\rangle$,            and
$a_{m;S,M,\alpha}$      is   the   sum  of    the  square of   the
(Clebsch--Gordan)        coefficients           of   the    states
$|m_1,m_2,m_3,m_4\rangle$, with $m_1 = m$  in the expansion of the
spin eigenstates    of the cluster in  terms of these determinants
\ref{equation:  Clebsch-Gordan  expansion}. For degenerate  states
it is important  to symmetrize  these  coefficients, giving  equal
weight to each of these states.

The total average magnetic moment per site is now:
\begin{equation}
\langle m \rangle = g\mu_B \sum_m m x_m.
\end{equation}
We define:
\begin{equation}
b_{m;S,M,\alpha} = a_{m;S,M,\alpha} - a_{m;0,0,\alpha_{0}},
\end{equation}
where the index $\alpha_{0}$ is  $(L,R)=(0,0)$. 
These $b$-coefficients are in table~\ref{table:b-coeff}.
We define:
\begin{equation}
B_{m;S,M,\alpha} = a_{m;S,M,\alpha} - a_{-m;S,M,\alpha}.
\end{equation}

The free energy of a system of $N_t$ clusters   on  a lattice  
can be written:
\begin{equation}
F = N_t\sum_{\gamma} [    \omega_{\gamma}E_{\gamma}        -  k_B T  \ln
W({\gamma})],
\end{equation}
where $E_{\gamma}   $ is the  energy  of  a cluster in state   $\gamma =
(S,M,\alpha)$,  and
$W({\gamma})$   is the number of ways  to distribute  a  set of states
$\{\gamma\}$, each occurring with probability $\omega_{\gamma}$   over the lattice.

In case  of the tetrahedron   as basis   cluster,  the    number of  ways  to
distribute these tetrahedron states over the lattice of CST  is calculated as
follows:  first distribute these tetrahedron states over  the whole  lattice.
$Z_t$ tetrahedra will  share  the  same  corner. So there are $N_t = Z_t/4 N$
tetrahedra, each  characterized   by a spin eigenstate    on the lattice. The
number of ways of putting these tetrahedron spin states on the lattice is:
\begin{equation}
Q = \frac{N_t !}{\prod_{\gamma}(N_t y_\gamma)!}
\end{equation}

In order that a certain distribution belongs to a realizable  one, the  local
spin moment for each  of these tetrahedra should  be  the same  on the shared
atoms, called  the overlap cluster.  In case of  the pyrochlore   lattice the
overlap cluster is the point cluster, and $Z_t = 2$.

This  restriction  on the distribution of these  tetrahedron  spin
states can not be applied rigorously, but  on the average: the CVA
for  the entropy.   At high temperatures the spins are distributed
over   all spin   eigenstates   of the tetrahedron:   this is  the
classical regime of spin  fluctuations.  At  very low temperature,
when the tetrahedron  spins are distributed over a small number of
tetrahedron  spin  states, one is in  the  regime of quantum  spin
fluctuations.   In  case  no classical   long--range    spin order
interferes this classical approximation of the  entropy  can break
down.  In case the  degeneracy of the tetrahedron  ground-state is
too low, the entropy   within  the CVA is negative below  a certain
temperature.     When,   at  low    temperature,     the tetrahedron
ground-state  is  a  single non  degenerate   singlet    state, no
correction is  necessary:   due to quantum  spin fluctuations  the
average  spin on each site vanishes: there is self--averaging. The
description of such a state is outside the validity regime of  the
CVA.  We take into account  correlated   spin  fluctuations of the
size of the tetrahedron, correlated spin  fluctuations over larger
distance are not described within the present  approach. With this
self--averaging   one might argue  that in a certain   temperature
range before  the system  enters the regime   of correlated  spin
fluctuations,  the state on  {\it nnn}  tetrahedra are independent
and one could approach   the susceptibility  by eq.~\ref{equation:
paramagnetic susceptibility}.  If that is the  case the asymptotic
Curie--Weiss temperature found from magnetic susceptibility   data
at relative low temperature  gives an exchange interaction 1/2 the
actual value.

The  correction within  the  CVA is  now  calculated   as follows.
Distributing   $N_t$ tetrahedra    over   the pyrochlore   lattice
generates $Z_t  N$ spins.  On  each site of the pyrochlore lattice
there   are $Z_t=2$ spins,  which  we call pseudo  sites.    First
calculate the number of ways of putting $Z_t N$  spins with at any
moment either spin  up or down on the  lattice  of $Z_t N$  pseudo
sites.  Of   these arrangements,  only a  fraction is correct: all
spin momenta  on the $Z_t$ pseudo sites belonging to the same site
are the  same.   Now the average    fraction    of  correct  point
configurations is:
\begin{equation}
\Gamma = \left[\frac{(N)!}{(x_{-1/2}N)!
(x_{1/2}N)!}\right]^{1-Z_t}.
\end{equation}
Combining this correction with  the number  of arrangements of the
tetrahedron configurations on the pyrochlore     lattice we obtain
for the average number of correct arrangements
\begin{equation}
W = (N!)^{1-3Z_t/4}
\left[ \frac{1}{\prod_{\gamma}(N y_{\gamma})!}\right]^{Z_t/4}
\left[\frac{1}{(x_{-1/2}N)!(x_{1/2}N)!}\right]^{1-Z_t}.
\end{equation}
In case of $s > 1/2$ one should replace the product $(x_{-1/2}N)!(x_{1/2}N)!
$ by the appropriate product for spin $s$: $\prod_{m=-s,s} (x_m N)!$.  

Substitution of this equation for the total number of spin  configurations in
the expression for the free energy and using  Stirling's formula   for  large
$N$ we obtain for the free energy per site
\begin{eqnarray}
f = F/N &=& Z_t/4 \sum_{\gamma} y_{\gamma} E_{\gamma} \nonumber\\
&+& k_B T \left[ Z_t/4
\sum_{\gamma} y_{\gamma}\ln(y_{\gamma})
+ (1-Z_t)\sum_{m=-s,s} x_m \ln(x_m)\right].\nonumber \\
&&
\end{eqnarray}

In order to obtain the equilibrium  values for  the probability  functions we
have   to minimize  the free  energy     with  respect    to  these functions
$y_{\gamma}$,   and find: 
\begin{eqnarray} 
0 = \frac{\partial f}{\partial  y_{\gamma}} =
Z_t/4 \left[(E_{\gamma}  - E_{0}) + k_B T \left(\ln\frac{y_{\gamma}}{y_{0}} +
\frac{4(1-Z_t)}{Z_t}       \sum_{m=-s,s}\frac{\partial       x_m   }{\partial
y_{\gamma}}\ln (x_m)\right)\right],\nonumber\\    && 
\end{eqnarray} 
where we used  the  fact  that  these distribution functions  $x$ and $y$ are
normalized.  $y_{0}$ and $E_{0}$   indicates the state with $S=0, L =  0, R =
0$.  Using  the definitions of the coefficients we can write: $\frac{\partial
x_m }{\partial y_{\gamma}} = b_{m;\gamma}$.  We eliminate  the state $0$ from
these equations and  write for the energy   of the eigenstates  in a magnetic
field:
\begin{equation}  
E_{S,M,\alpha}      = E^{0}_{S,M,\alpha}   + g\mu_B  M H.   
\end{equation}  
where $E^{0}_{S,M,\alpha} = E^{0}_{S,-M,\alpha}$.  The equilibrium  equations
become: 
\begin{equation}           
\frac{y_{S,M,\alpha}}{y_{S,-M,\alpha}}=\prod_m x_m^{2C_t\,B_{m;S,M,\alpha}}   \exp(2M h/T),  
\end{equation}  
where we defined: $h = g\mu_B H$,  and we set $k_B = 1$ and the constant $C_t
= 2(Z_t-1)/Z_t$. 

In case the spin  system   is non--magnetic -- no long--range order -- one can
easily  solve these  equations. The distribution functions are independent of
the  magnetic quantum number  $M$, and the  probability that a spin is in one
of its $2s+1$   magnetic moment states $m$ is in  case of  Heisenberg  spins:
$1/(2s+1)$.  

In case  of   finite single--ion anisotropy,  the  $x_m$   should   be
evaluated  using the  appropriate   probability distributions  for the
non--magnetic solution.  

The distribution over the tetrahedron states is in the
non--magnetic case without magnetic field given by:
\begin{equation}
y_{S,M,\alpha}  = y_0  \exp(-(E_{S,M,\alpha}  - E_0)/k_B T),
\end{equation}
and 
\begin{equation}
y_0 = 1 - \sum_{S,M\alpha \ne 0} y_{S,M,\alpha}.
\end{equation}

\subsubsection{The entropy $\cal{S}$ for $s=1/2$.\label{subsection:-entropy}}

The  free energy    is for $s=1/2$   in the disorder    regime:
\begin{equation}  
F  =
-\frac{1}{2}k_B  T \ln Z_0 + k_B  T \ln 2 
\end{equation}   and  the
entropy in this regime is:
\begin{equation}   
{\cal{S}}/k_B = \frac{1}{2}\ln
(Z_0/4)  +  \frac{1}{2}T   \frac{d\; \ln Z_0}{d\;  T} 
\end{equation} 
where
\begin{equation} Z_0 =
\sum_{\gamma}\exp   (-E_{\gamma}/k_B    T)  
\end{equation} 

In  the disorder     regime   the single--site    spin    momentum
probabilities       $x_{m=\pm      1/2}=1/2$   are constant.   For
$T\rightarrow  0$ the   last contribution to   the entropy can  be
neglected.   We  find  that for $Z_0  <  4$ the  entropy   becomes
negative:  the  restrictions  imposed   by the CVA are too strong.
The temperature where this occurs we call $T_S$.  We find  that in
case there  is  no long--range order  at any finite   temperature,
then in case the tetrahedron has a ground state  with a degeneracy
less than  4, the spin system has  in some way to enter a state of
correlated  spin motion  in  order to   satisfy the  thermodynamic
requirement  that  ${\cal{S}}  \ge  0$.  Already far above $T_S$ one
should observe  strongly  correlated   spin fluctuations extending
over more than one tetrahedron. 

It  seems that   the  restrictions  to  determine   the number  of
realizable configurations   imposed by the CVA  are too strong. At
low temperatures,    when  only   singlet  tetrahedron  states  are
occupied,   there is self--averaging and no CVA--type restrictions
are necessary.  

\subsubsection{The magnetic
susceptibility.\label{subsection:-susceptibility}}

In order  to  calculate the magnetic  susceptibility  we linearize
the functions  which  describe  the   distribution   of   the spin
configurations in the magnetic   field $H$ and  the  magnetization
${\cal{M}}$ as follows. First we write:
\begin{equation}
y_{S,\pm M,\alpha} = y_{0;S,M,\alpha}(1 \pm m_{S,M,\alpha}),
\end{equation}
where $m_{S,M,\alpha}$ is a new function linear in 
the magnetization and
magnetic field. In the following the summations over $M$ 
are restricted to $M \ge 0$. 
We linearize the exponent which contains the magnetic field and get:
\begin{equation}
(1 + 2m_{S,M,\alpha})= (1 + 2Mh)  \prod_{m=-s,s}  x_m^{-2C_t\,B_{m;S,M,\alpha}}. 
\end{equation}
Next we write an expression for the single--site distributions  in
terms of these cluster distributions, and we find
\begin{equation}
x_m = x_{0,s,m} + \sum_{S,M > 0,\alpha  \ne 0} 
B_{m ;S,M,\alpha}\,y_{S,M,\alpha}\,m_{S,M,\alpha},
\end{equation}
where we defined 
\begin{equation}
 x_{0,s,m} =\sum_{S,M,\alpha  \ne 0}b_{m ; S,M,\alpha}y_{S,M,\alpha} + b_m(0). 
\end{equation}
We have now reduced the number of unknown state distribution functions to the
states without  taking into account the $M$ degeneracy. For $s=1/2$ there are 6. 

One can easily see that $B_m = -B_{-m}$. So we can write ($h =
H/k_B T$):
\begin{eqnarray}
1 + 2m_{S,M,\alpha_S}& = &(1 + 2Mh) \nonumber \\
&&
\left[\frac{\prod_{m> 0 } 
\left[x_{0s} + \sum_{S,M > 0,\alpha  \ne 0} B_{m; S,M,\alpha}
y_{S,M,\alpha}m_{S,M,\alpha_S})\right]}{\prod_{m>0} 
\left[x_{0s} - \sum_{S,M > 0,\alpha  \ne 0} B_{m; S,M,\alpha} 
y_{S,M,\alpha}m_{S, M,\alpha})\right]}
\right]^{-2C_t \; B_{m; S,M,\alpha}}.
\end{eqnarray}
This can now easily be evaluated to lowest order in the magnetic field and
the magnetization for each $m$ state.
We define the following quantity:
\begin{equation}
m_m = \sum_{S,M > 0,\alpha  \ne 0}
B_{m;S,M,\alpha}y_{S,M,\alpha}m_{S,M,\alpha},
\label{equation:definition-m_m}
\end{equation}
and we can now solve for $m_m$:
\begin{eqnarray}
m_m &=& K_m h - C_t \frac{2}{x_{0s}}Y_m m_m - 
C_t \frac{2}{x_{0s}}\sum_{m_1\ne m;m_1>0}Y_{m m_1}m_{m_1}, 
\label{equation:basic-tetrahedron}
\end{eqnarray}
where we defined the following functions:
\begin{eqnarray}
K_m &=& \sum_{S\ne 0,M>0,\alpha} B_{m ;S,M,\alpha}M y_{S,M,\alpha},\\
Y_{m_1 m_2} & = & \sum_{S\ne 0,M>0,\alpha} 
B_{m_1 ; S,M,\alpha}B_{m_2 ; S,M,\alpha} y_{S,M,\alpha}.
\end{eqnarray}

In   the cases $s=1/2$   and $s=1$ this set of  equations   can be
solved easily.  In the case $s=1/2$ one finds
\begin{equation}
m_{1/2} = K_{1/2}/(1 + C_t \frac{2}{x_{0,1/2}} Y_{1/2})h,
\end{equation}
with $x_{0,1/2} = 1/2$.

The  magnetization          can         be     written         as:
\begin{equation}
{\cal M} = \sum_m m x_m = 2 \sum_m m m_m
\end{equation}
The second equation for ${\cal M} $ follows easily from the definition of
$x_m$ and $m_m$.
The factor 2 appears, because the sum in $m_m$ over $M$ is only for $M>0$.

The magnetic susceptibility is now given by: 
\begin{equation}
\chi = {\cal M} /H = \frac{2 \sum_m m m_m}{H}
\end{equation}
and 
in case of spin $s=1/2$ we get
\begin{equation}
\chi = \frac{m_{1/2}}{H} = \frac{K_{1/2}}{k_B T(1 + 4 C_t Y_{1/2})}
\end{equation}
The functions $K$  and  $Y$ can  easily  be calculated   from  the
Clebsch--Gordan    coefficients and the probability   distribution
functions   for the non--magnetic     system.    We  find that the
asymptotic Curie--Weiss temperature  in the CVA is equal to the MF
result.

For $s=1/2$ we find for the susceptibility in case of zero uniaxial 
fields ($D_T =\Delta = 0)$:
\begin{equation}
\chi = \frac{g^2\mu_B^2}{ T}\frac{5e^{6J/T}+3e^{2J/T}}{2+6e^{2J/T}}.
\end{equation}
where we have set $k_B = 1$.
A comparison of the CVA result and an exact high--temperature expansion
shows that the CVA 
\begin{equation}
\chi =\frac{g^2\mu_B^2}{ T}\left[1 + 3x + 6x^2+\frac{17}{2}x^3\cdots\right] 
\end{equation}
already deviates at second order in $x = J/ T$~\cite{1D-AfH-s=1/2-alternating}
\begin{equation}
\chi = \frac{g^2\mu_B^2}{ T}\left[1 + 3x + 9x^2 + \cdots\right].
\end{equation}
We  are not aware of any high--temperature   expansion for the CST
lattice, only for the Kagome~\cite{CAF-HTE-KAgome}.

The magnetization is given by the solution of the following implicit
equation in ${\cal{M}}$:
\begin{equation}
{\cal{M}} =
\frac{(A_{1,2}^- + \frac{1}{2}A_{\frac{1}{2},1}^-)
e^{\frac{6(J+D_T)+4\Delta}{T}} +
A_{\frac{1}{2},1}^-
e^{2\frac{J+D_T}{T}}\left(e^{2\frac{\Delta}{T}} + 
\frac{1}{2}e^{4\frac{\Delta}{T}}\right)
}
{1+e^{4\frac{\Delta}{T}} +
(A_{1,2}^+ +  A_{\frac{1}{2},1}^+ + 1)e^{\frac{6(J+D)+4\Delta}{T}} +
(A_{\frac{1}{2},1}^+ + 1)
e^{2\frac{J+D_T}{T}}
\left(2 e^{2\frac{\Delta}{T}} +  
        e^{4\frac{\Delta}{T}} \right)
}
\label{equation:-magnetization-s=1/2}
\end{equation}
where we defined: $x = (1+{\cal{M}})/(1-{\cal{M}})$, $A_{1,2}^{\pm} = (x e^{2H/T}\pm
\frac{1}{x} e^{-2H/T})e^{-12D_T/T}$, and  $A_{\frac{1}{2},2}^{\pm} =
(\sqrt{x} e^{H/T}\pm
\frac{1}{\sqrt{x}} e^{-H/T})e^{-3D_T/T}$.

\sectie{RESULTS AND DISCUSSION.\label{section:-results-discussion}}

\subsection{The critical temperature for Heisenberg spins.
  \label{subsection:-Tc}}   

We determine  the critical    temperature as the temperature where
the susceptibility    diverges.    The    results    are  in table
\ref{table:-critical-temperature-tetrahedra}.     In  case $s=1/2$
and only   Heisenberg     interactions,      there  is no magnetic
long--range   order  at any  finite temperature, neither for F nor
for AF interactions.   In  our CVA to  the pyrochlore lattice,  AF
Heisenberg  spin  interactions   never give  classical long--range
order.   In case of ferromagnetic Heisenberg interactions the  CVA
values  are about 60 \% of the usual  Mean Field  values, slightly
increasing -- as expected -- for larger spin.  

We calculated for  $s=3/2$     the  critical   temperature  for  F
Heisenberg        exchange         from      a   high--temperature
approximation~\cite{HTE-ZnxCd1-xCr2Se4} with  6 coefficients using
the Pad\'{e} extrapolation and find  $T_c/J =  -7.892$.   Compared
with this,   in  principle   exact calculation,   the CVA critical
temperature  is still too high, indicating   that spin fluctuations
are  stronger  than accounted for in  our application  of the CVA.
Note  also that  our critical  temperatures    for   ferromagnetic
Heisenberg  interactions are  appreciably lower than those for the
same coordination   number  using   the interpolation  equation of
Rushebrook                    and                             Wood
\cite{Crit-temp-interpolated-Rushbrooke-Wood}.  

We studied    the effect of increasing  the number   of tetrahedra
sharing   the same  corner.  We found that  any  increase  however
small will lead to a finite critical  temperature   for $s=1/2$ in
case of ferromagnetic  Heisenberg exchange. This is illustrated in
figure
\ref{figure:-critical-temperature-ferromagnetic-order-coordination-number}.
A  similar conclusion   was     reached   by    Pinettes        et
al~\cite{CHAF-Pyr-Entropy-Energy}   who -- using MC simulations --
analyzed  the ordering temperature  going from  a pyrochlore  to a
fcc structure.   They   found that for very small values  of   the
ratio of the extra exchange  interaction   -- needed  to transform
the pyrochlore lattice into the  fcc lattice  --  and the exchange
interaction   in  the pure pyrochlore lattice, the  system shows a
collinear long--range order. 

\subsection{Ground state Phase Diagram for
$s=1/2$.\label{subsection:-GS_s=1_2}}

In order to  derive the ground  state phase diagram for $s=1/2$ we
calculated   the energy  levels  of  the   spin     states  of the
tetrahedron,      which        can       be      found    in table
\ref{table:-energy-levels-s=1/2}   and  are illustrated  in figure
\ref{figure:-energy-levels-s=1/2}.     The phase diagrams   in the
$d_T$--$\delta$   plane, where $d_T = D_T/J$  and $\delta_T    =
\Delta_T/J$, are   in figure  \ref{figure:-phase-diagrams}    for
ferromagnetic as well  as antiferromagnetic interactions.  We note
that  for various values of the distortion  the   ground  state is
non-magnetic.   The lines  separating  the  various   regions  can
easily be found from the expressions for the energy.

In case of F interactions we find the following equations  for the
boundaries     between        the various ground  states:       \\
($S=0,M=0$)--($S=2,M=0$):         $2\delta       +   3d_T   +3 = 0$;\\
($S=0,M=0$)--($S=2,M=2$):       $2\delta       -   3d_T  +3  =   0$;\\
($S=2,M=2$)--($S=2,M=0$):    $d_T  = 0$.\\ 
So we find 3 regions:  one region where the spins show long--range
order  ($S=2,M=2$),  and two regions  where  we find  a non magnetic
ground state: one where  the   ground state of the spin system  is
build  from singlet   dimers ($S=0$,$M=0$,$L=0$,$R=0$),  and   one
build from the nonmagnetic $M=0$ component of the quintet state,
indicated by RVT2. 

In case of AF interactions we find the following boundary equations:\\
(S=2,M=2)--(S=0,M=0): $2\delta-3d_T+3=0$; \\
(S=0,M=0)--(S=2,M=0): $2\delta+3d_T+3=0$; \\
(S=0,M=0)--(S=2,M=0): $d_T = -1$; \\
(S=2,M=2)--(S=0,M=0): $d_T = 1$.\\

In this case we  find  four  regions: the region with  a classical
long--range   AF  order,  and three   regions without     magnetic
long--range  order, the region indicated  by RVB  in the  diagram,
with  a ground   state  build up from singlet    dimers,  a region
characterized    by  a  singlet   tetrahedron   ground  state with
compensating  triplet spins $L=1$, $R=1$, indicated by a Resonance
Valence Tetrahedron  state: RVT1,  and  a region where  the ground
state is build  from the nonmagnetic  spin  moment   $M=0$  of the
tetrahedron quintet state, indicated by RVT2.

In case  of AF   interactions    and $d_T >   0$, the ground-state
becomes magnetic for $d_T > d_{T,c}=(4s+1)/(8s+1)$.   In  case  of
$s=1/2$ the critical  value  $d_{T,c} = 1$ and for very large $s$,
$d_{T,c}\rightarrow 1/2$.  If we translate this to  the  case that
dipole--dipole    interactions    are the source of the tetragonal
anisotropy, then  $D_T  = D_D/8$, and  in  the  large $s$ limit we
find   for the critical   ratio   of  the $nn$ exchange    and the
dipole--dipole interaction:  $J/D_D = 1/4$. For smaller ratios the
$S=2,M=2$ state drops below the singlet  $S=0,M=0$ state.  Whether
this can be applied to the  Ising systems   discussed  in a recent
review  by Bramwell  and  Gingras  \cite{review-spin-ice},  is not
clear.  In  case we make the  necessary corrections for $J$; their
exchange interaction  is -1.5 times  our exchange constant, and we
find from their  work the following condition for magnetic  order:
$J/D < 1$ , which   is of the  same  order of magnitude as we have
derived. 

The boundary   separating     the magnetic  from the non--magnetic
solutions is not a sharp line: these two regions are separated  by
a re-entrant  region  where there  are two critical  temperatures.
The extent   of  this region increases with increasing tetrahedron
distortion         like       illustrated           in      figure
\ref{figure:-reentrant-region-ferromagnetic-nn-interactions}.   In
figure   \ref{figure:-critical-temperature-tetrahedron-anisotropy}
we give the critical temperature $k_B T_c /J  $ as  a function  of
the tetrahedral  anisotropy   $d_T$,   for various  values  of the
exchange anisotropy $\delta$.  This re-entrant behaviour is due to
the occupation   of a magnetic   state with increasing temperature
i.e.  an    entropic      effect.         It      is      a   type
order--by--disorder~\cite{order-by-disorder-Villain}   transition:
an increase in thermal spin fluctuations induces magnetic order.

Koga and  Kawakami~\cite{Frus-HAF-Pyr}   have also derived a phase
diagram for  the  Quantum Heisenberg  pyrochlore system    with AF
interactions.  They   studied  the    influence   of  the exchange
anisotropy $\Delta$.  When comparing  their   result with ours  we
might  identify    the plaquette state with  our $S=0,M=0,L=1,R=1$
state and their dimer state with our  $S=0,M=0,L=0,R=0$ state. For
$\delta = 0$, for an undistorted tetrahedron these two states  are
degenerate.     Whether    the system   with  $\Delta=-J$   and AF
interactions,  i.e.,  the  total $L$ and $R$  exchange interactions
vanish, should show magnetic  order  is questionable. It is a very
open 3D network with a very low coordination number of 4.

In            figure
\ref{figure:-critical-temperature-tetrahedron-anisotropy}       we
compare the ferromagnetic  order temperature $T_c/J$ as a function
of tetragonal anisotropy $d_T = D_T/J$ for various  values of  the
exchange  anisotropy $\delta  = \Delta/J$.    We  find that $T_S <
T_c$ in case of finite $T_c$.

In  the regions of the phase diagram where the  ground state has a
degeneracy smaller  than 4, there  is  a finite temperature  $T_S$
where the    entropy   in  the  CVA   becomes     negative.     In
figure~\ref{figure:-entropy-ferro-s=1/2}      we present $W$  -the
number  of  states  accessible   by the  spin  system so that  the
entropy per  site is ${\cal{S}}/k_B  = \frac{1}{4}\ln W$ - for the
case of ferromagnetic  exchange $J=-1$  as a function  of $T/|J|$,
without  any  perturbations.    Because  the spin  system does not
order for  $T>0$, the   entropy  remains finite for all $T>0$.  At
high temperatures   the  entropy approaches    the   value   for a
paramagnetic spin system ($\ln 2$).

We surmise that  below  $T_S$ the spin system   enters  in a state
with strongly correlated  quantum  spin fluctuations.   The latter
takes over from the slowed  down thermal fluctuations. This occurs
in all regions without magnetic moment degeneracy ($M=0$). 

Such a ground  state degeneracy can also be lowered by a  magnetic
field.   In figure~\ref{figure:-entropy-magnetic-field-s=1/2}   we
present some  results of $T_S$  as  a  function of the  tetragonal
anisotropy for $\delta =0$ and $\delta = -3$.   In the first  case
the magnetic  field  favours the  state    with   extended    spin
fluctuations  for all $d_T$'s while  in case of $\delta=-3$  these
fluctuations   are  suppressed  with increasing   field  for small
tetragonal anisotropy, while   for   large  $d_T$     the  quantum
correlated spin fluctuations are favoured by a magnetic field.

We  have  determined  the extend of the re--entrant   region   for
$s=1/2$                 to                   7/2.               In
fig.~\ref{figure:-crit-temp-single-ion-anis-s=3/2}
we present the re--entrant   region for $\delta =$ for $s=3/2$. In
table~\ref{table:-reentrant}      we   give $T_{max}$, $D_{T,max}$
values  which  characterize  the extend of the re--entrant  region
for $s=1$  to  $s=7/2$: this re--entrant   region increases   with
increasing  spin. We expect to find  similar behaviour in case  of
trigonal single--ion anisotropy. 

\subsection{Magnetization ${\cal{M}}$  as a function of field and 
temperature: the F case.\label{subsection:-M-T-H-Ferro}}

In figure~\ref{fig:-ferro-M-H}   we  present   the magnetization  as a
function of   the magnetic field for ferromagnetic  exchange  ($J=-1$)
and  $s=1/2$ along the lines A-C ($\delta  = -3$ and B-D  $\delta = 1$
in  figure~\ref{figure:-phase-diagrams},      for $T  = 0.01|J|, 1|J|,
5|J|$.   We  can distinguish  three  types  of  behaviour,   which  is
directly related to the ground state in the phase diagram.

\begin{enumerate}
\item
In the  region  of (ferromagnetic) order    ($S=2$,    $M=2$)  the
magnetization   behaves  in the usual  way as   a function of  the
magnetic field.   We note that for the unperturbed case ($\delta =
0$, $d_T=0$)   only a very  small  magnetic field is  necessary to
magnetize the system.
\item
In the region characterized by ($S=0$, $M=0$,  $L=0$, $R=0$), that
is the region of the  phase diagram where due to  the $\delta$ and
$d_T$ anisotropy,  the magnetic states are pushed far above the non
magnetic ground state.  Here we observe, at very low  temperatures
as a  function of the  magnetic field, a single first--order phase
transition  from a non--magnetic  state to a completely    ordered
state: $ {\cal{M}}={\cal{M}}_{sat}$      for  a field  $h_c = -3-2
\delta  + 3d_T$.   This transition   smears out  with   increasing
temperatures.  At  $T=0$ this is what is  called a Quantum   Phase
Transition.
\item
The region characterized  by ($S=2$,  $M=0$,  $L=1$,   $R=1$).  In
this region the   magnetization   as a function of the field shows
two first--order phase   transitions as a function of the field at
very  low temperature:   one at $h_{c1}  = 3d_T$   and the next at
$h_{c2}  =  9d_T$ (Quantum Phase Transitions).  The plateau in the
magnetization  is at ${ \cal{M}}= 1/2  {\cal{M}}_{sat}$.   Like in
the previous case   these   transitions    are smeared   out  with
increasing temperature. 
\end{enumerate}
Note that in various  cases   the magnetization  increases  nearly
linear  with the field, and in some case it changes from  a convex
to a concave function with increasing field.

These magnetic  field dependences      lead to a rather   peculiar
temperature       dependence       of   the   magnetization.    In
figure~\ref{fig:-ferro-M-H} we present the temperature  dependence
of the  magnetization  along  the  B-D line ($\delta = 1$) in  the
ferromagnetic phase diagram,  for  positive  $d_T$ values  ($S=2$,
$M=0$, $L=1$, $R=1$).  Fundamentally   we observe three types   of
behaviour  of the  magnetization as a function of the temperature.
For small field the magnetization  has  a maximum as a function of
the temperature for all values  of $d_T>0$.  With increasing field
for small    values      of   $d_T$  the system is ordered  at low
temperatures,  while for large values  of $d_T$  the magnetization
has  a  maximum.     For   intermediate     values   of $d_T$ (see
figure~\ref{fig:-ferro-M-H})   the magnetization as  a function of
temperature  exhibits a relative minimum.  Note  that in case this
relative  minimum is present in the magnetization as a function of
temperature,        than   the zero  temperature    magnetization:
${\cal{M}}(T\rightarrow       0)\rightarrow   0.5{\cal{M}}_{sat}$,
independent of the applied magnetic  field.  This is related  with
the plateau ${\cal{M}}=1/2{\cal{M}}_{sat}$    in the magnetization
as a function of field.

In the region ($S=0$,  $M=0$,   $L=0$,   $R=0$) one only finds one
peak in the magnetization as a function of temperature.

In figure~\ref{figure:-reentrant-M-T-s=1/2}       we present   the
magnetization   in a small field for  $\delta = -3$  for   various
values of the  tetragonal  anisotropy $d_T$  in  the   re--entrant
region.   The peak in the magnetization  is clearly visible.  Note
that  even    outside     the ordered region  there   is  a finite
magnetization which    does not  decrease  to  zero  for vanishing
field.

In figure~\ref{fig:-specific-heat-reentrant}    we illustrate  the
behaviour   of the specific heat as a function  of temerpature  in
the re--entrant  region for ferromagnetic   exchange.   Note  that
there appear  two peaks on top of a Schottky  peak.  The behaviour
of the specific   heat at low temperatures is uncertain as the CVA
breaks  down   (negative entropy).    We conjecture  that trigonal
perturbations can induce similar beaviour.

\sectie{SUMMARY AND CONCLUSION\label{section:-summary-conclusions}}

We conclude   from our calculations    that in case of AF {\it nn}
exchange only there is no  classical long--range spin order on the
pyrochlore  or B sublattice of the spinel structure  at any finite
temperature.   This  is a consequence     of  the high  degree  of
frustration on this  lattice  for AF  interactions     and its low
coodination number.  In case  of  the  simple cubic  lattice, which
has the same coordination number  (6) as the pyrochlore   lattice,
there is spin  order below a certain  finite temperature with {\it
nn} interactions only.

In case  of F {\it nn} exchange  only there  is  no spin order for
finite temperatures  for $s=1/2$.  For $s>1/2$  one finds a finite
critical temperature which is appreciably  smaller than that found
in the MF approach.  However from  a HTE using the coefficients in
~\cite{HTE-ZnCdCr2Se4-2}   one  finds  an even smaller    ordering
temperature,  which indicates  that even in the case of F {\it nn}
Heisenberg--type interactions,   spin order is frustrated.  Such a
frustration of spin   order  in  case of  F interactions  has been
discussed          by              Bramwell                    and
Harris~\cite{Spin-Ice-Ising,Spin-Ice-Ho2Ti2O7}     for the case of
classical Ising spins,  each pointing  along the local $<\!111\!>$
axis  of  a tetrahedron:   the 2 spin   in 2 spin out ground state
configuration  is sixfold   degenerate,  leading  to a finite zero
temperature  entropy.  In case of  quantum spins there is no  such
degeneracy.  The reason  why the $s=1/2$  pyrochlore system with F
interactions does  not order,    is due  to  the low  coordination
number (6).  One could also argue that  from the  point of view of
the $S=2$ tetrahedron  spin, the coordination number    is only 4,
and the smallest closed loop is a hexagon.   Heisenberg   spins on
such a four--fold coordinated structure will probably not order.

We did not include  the  effects   of  long--range  dipole--dipole
interactions, which recently have been proposed  to be responsible
for the so--called  dipolar  spin--ice ground state.  Our appraoch
is valid in  case $J_{nn} > D_D$ \ref{eq:-dipole-dipole},   so for
systesm containing  magnetic    transition  metal ions like spinel
with   magnetic  ions   on the    B   sublattice    ZnCr$_2$O$_4$,
CsNiFeF$_6$,                               CsNiCrF$_6$,
CsMnFeF$_6$\cite{review-experimental-Harris-Zinkin},
GeCu$_2$O$_4$\cite{CuGe2O4-spinel-1D},
LiV$_2$O$_4$~\cite{LiV2O4-Pyrochlore},       than to  systems with
rare--earth iopns on a pyrochlore lattice.

We have found  in  various regions of parameter  space re--entrant
magnetic  behaviour when lowering the  temperature:  non--magnetic
(Paramagnetic)    $\rightarrow$    Ferromagnetic     $\rightarrow$
non--magnetic (Two-Dimer-Singlet  or Tetrahedron   Singlet).   The
high temperature phase is paramagnetic, while the low  temperature
phase is some type  of spin fluctuating  state.   Such re--entrant
behaviour has  been reported  as a function  of the magnetic field
for the  garnet  Gd$_3$Ga$_5$O$_{12}$    by  \cite{review-Ramirez,
review-Schieffer-Ramirez,    GGG-Petrenko-1, GGG-Petrenko-2}   and
recently been discussed by Tsui et al\cite{Reentrant-Mag-SL}.

The magnetization as a function of the  field depends  strongly on
the model   parameters.  We find magnetization curves which show a
single step for large field  or  a  two step  behaviour.  Stepwise
behaviour has  been  found  for classical  Heisenberg  magnets  by
Zhitomirsky et  al~\cite{Field-induced-order}.     See  also   the
discussion   of  magnetization    plateaus     by  Lhuillier   and
Misguich~\cite{review-frus-magn}.   Such steps have  been observed
in      various                          pyochlore
systems\cite{review-experimental-Harris-Zinkin,
review-Magnetization-RE2Ti2O7}. 

We have also calculated  the  specific  heat for $s=1/2$.  Details
of the  calculation   of the  specific   heat  will   be published
elsewhere.    For  the zero-field  specific heat  of non--magnetic
systems we find usually  a single  peaked  Schottky--like anomaly,
and in case  of  the   re-entrant  phase we find two  peaks in the
specific heat, possibly   on top of a Schottky--like peak.    Such
peaks  have recently  been observed  in  the   specific    heat of
Gd$_2$Ti$_2$O$_7$~\cite{Gd2Ti2O7-Mag-Field-Ind-Phase-Tran}.  

Summarizing  we find  indications from this CVA of a rich magnetic
behaviour for spins on the highly  frustrated  pyrochlore  lattice
as  a function of tetragonal  distortions:   Two dimer singlet  or
tetrahedron ground states, re--entrant  behaviour  due to entropic
ground  state   selection, magnetic   field induced order  Quantum
Phase Transitions),  together with steps in the magnetization as a
function of field, Schottky like anomalies in the specific heat. 

\section*{Acknowledgments}     We acknowledge    a grant from CNPq
(300928/97--0).  We also acknowledge  Prof.  Luciano  Peixoto  for
critically reding the manuscript. 
\newpage

%\bibliography{Pyrochlore}

%\bibliography{Pyrochlore}

\begin{thebibliography}{10}

\bibitem{Lacorre}
P.~Lacorre.
%\newblock The constraint functions: an attempt to evaluate the constraint rate
%  inside structures that undergo ordered magnetic transitions.
\newblock {\em J. Phys. C: Solid State Phys.}, {\bf 20 }:L775--L781, (1987).

\bibitem{Anderson-RVB-triangles}
P.W. Anderson.
\newblock {\em Mat. Res. Bull.}, {\bf 8}:153, (1973).

\bibitem{Fazekas-Anderson-1974}
P.~Fazekas and P.~W. Anderson.
\newblock {\em Phil. Mag.}, {\bf 30}:23, (1974).

\bibitem{review-Ramirez}
A.~P. Ramirez.
\newblock {\em Strongly geometrically frustrated magnets}.
\newblock {\em Annu. Rev. Mater. Sci.}, {\bf 24}:453, (1994).

\bibitem{review-Schieffer-Ramirez}
P.~Schiffer and A.~P. Ramirez.
%\newblock Recent experimental pregress in the study of geometrical magnetic
%  frustration.
\newblock {\em Comments Cond. Mat. Phys.}, {\bf 18}:21, (1996).

\bibitem{Review-Geo-Frus-Moessner}
R.~Moessner.
%\newblock Magnets with strong geometric frustration.
\newblock {\em Can. J. Phys.}, {\bf 79}:1283--1294, (2001).

\bibitem{Loc-latt-disorder-frus-SG-Y2Mo2O7}
C.~H. Booth, J.~S. Gardner, G.~H. Kwei, R.~H. Heffner, F.~Bridges, and M.~A.
  Subramanian.
%\newblock Local lattice disorder in the geometrically-frustrated spin glass
%  pyrochlore y2mo2o7.
\newblock {\em Phys. Rev. B}, {\bf 62}:R755, (2000).

\bibitem{ZnCr2O4-QAF}
H.~Martinho, N.~O. Moreno, J.~A. Sanjurjo, C.~Rettori, A.~J. Garcia-Adeva,
  D.~L. Huber, S.~B. Oseroff, W.~Ratcliff, S.-W. Cheong, P.~G. Pagliuso, J.~L.
  Sarrao, and G.~B. Martins.
%\newblock Magnetic properties of the frustrated afm spinel zncr2o4 and the spin
%  glass zn1-xcdxcr2o4 (x=0.05;0.10).
\newblock {\em Phys. Rev. B}, {\bf 64 }:24408, (2001).

\bibitem{SG-Ferro}
S.~Franz, M.~Mezard, Ricci-Tersenghi, M.~Weigt, and R.~Zecchina.
%\newblock A ferromagnet with a glass transition cond-mat/0103026.
\newblock {\em Europhys. Lett.}, {\bf 55 }:465, (2001).

\bibitem{Geo-frus-Spin-Ice-neg-exp}
A.~P. Ramirez, C.~L. Broholm, R.~J. Cava, and G.~R. Kowach.
%\newblock Geometrical frustration, spin ice and negatie thermal expansion --
%  the physics of underconstraint.
\newblock {\em Physica B}, {\bf 280}:290--295, (2000).

\bibitem{Pyr-JT-Spin}
Y.~Yamashita and K.~Ueda.
%\newblock Spin driven jahn-teller distortion in a pyrochlore system.
\newblock {\em Phys. Rev. Lett.}, {\bf }:to appear.

\bibitem{order-by-distortion}
O.~Tchernyshyov, R.~Moessner, and S.~L. Sondhi.
%\newblock Order by distortion and string modes in pyrochlore af.
\newblock {\em Phys. Rev. Lett.}, {\bf 88 }:67203, (2002).

\bibitem{order-by-disorder-bond-disorder}
L.~Bellier-Castella, M.J.P. Gingras, P.C.W. Holdsworth, and R.Moessner.
%\newblock Frustrated order by disorder: the pyrochlore af with bond disorder.
\newblock {\em Can. J. Phys.}, {\bf 79}:1365--1371, (2001).

\bibitem{CF-Mag-Field-Liquid-Gas}
M.~J. Harris, S.~T. Bramwell, P.~C.~W. Holdsworth, and J.~D.~M. Champion.
%\newblock Liquid-gass critical behaviour in a frustrated pyrochlore
%  ferromagnet.
\newblock {\em Phys. Rev. Lett}, {\bf 81}:4496--4499, (1998).

\bibitem{Dy2Ti2O7-GS-Magnetic-Field}
K.~Matsuhira, Z.~Hiroi, T.~Tayama, S.~Takaga, and T.~Sakakibara.
%\newblock A new macroscopic ground state in spin ice compiound dy2ti2o7 under a
%  magentic field.
\newblock {\em J. of Phys.: Cond. Mat.}, {\bf 14 }:L559, (2002).

\bibitem{Reentrant-Mag-SL}
Y.~K. Tsui, J.~Snyder, and P.~Schiffer.
%\newblock Thermodynamic study of excitations in a 3d spin liquid.
\newblock {\em Phys. Rev. B}, {\bf 64 }:12412, (2001).

\bibitem{order-by-disorder-Villain}
J.~Villain, R.~Bidaux, J.~P. Carton, and R.~J. Conte\'{e}.
%\newblock Order as an effect of disorder.
\newblock {\em J. Phys. (Paris)}, {\bf 41}:1263., (1980).

\bibitem{Villain-o-b-d}
J.~Villain.
%\newblock {Insulating Spin Glasses }.
\newblock {\em Z. Phys. B}, 33:21, (1979).

\bibitem{Harm-SW-QAF-pyr}
R.~R. Sobral and C.~Lacroix.
%\newblock Order by disorder in pyrochlore af.
\newblock {\em Solid State Comm.}, {\bf 103}:407--409, (1997).

\bibitem{Crit-Prop-Pyr-AF}
J.~N. Reimers, J.~E. Greedan, and M.~Bj\"{o}rgvinsson.
%\newblock Critical properties in highly fruistrated pyrochlore
%  antiferromagnets.".
\newblock {\em Phys. Rev. B}, {\bf 45}:7295--7306, (1992).

\bibitem{Critical-Dynamics-Frustration}
V.~N. Kotov, M.~E. Zhitomirsky, and O.~P. Sushkov.
%\newblock Critical dynamics of singlet excitations in a frustrated spin system.
\newblock {\em Phys. Rev. B}, {\bf 63}:64412, (2001).

\bibitem{Spin-Ice-Ising}
S.~T. Bramwell and M.~J. Harris.
%\newblock Frustration in ising-type spin models on the pyrochlore lattice.
\newblock {\em J. Phys.: condens. matter}, {\bf 10}:1215--1220, (1998).

\bibitem{Ice-Rules-Pyr}
R.~Siddharthan, B.~S. Shastry, A.~P. Ramirez, A.~Hayashi, R.~J. Cava, and
  S.~Rosenkranz.
%\newblock Ising pyrochlore magnets: low temperature prop., ice rules and
%  beyond.
\newblock {\em Phys. Rev. Lett.}, {\bf 83}:1854, (1999).

\bibitem{Dipolar-Spin-Ice-Numerical}
R.~G. Melko, B.~C. den Hertog, and M.~J.~P. Gingras.
%\newblock Long range order at low temperatures in dipolar spin ice.
\newblock {\em Phys. Rev. Lett.}, {\bf 87}:67203, (2001).

\bibitem{Dipolar-Order-Geo-Frus-AF}
S.~E. Palmer and J.~T. Chalker.
%\newblock Order induced by dipolar interactions in a geometrically frustrated
%  af.
\newblock {\em Phys. Rev. B}, {\bf 62}:488, (2000).

\bibitem{Dipolar-Origin-Spin-Ice}
B.~C. den Hertog and M.~J.~P. Gingras.
%\newblock Dipolar interactions and origin of spin ice in ising pyrochlore
%  magnets.
\newblock {\em Phys. Rev. Lett.}, {\bf 84}:3430, (2000).

\bibitem{Dipolar-Origin-Spin-Ice-MFT}
M.~J.P. Gingras and Byron~C. den Hertog.
%\newblock Origin of spin ice behaviour in ising pyrochlore magnets with long
%  range dipole interactions: an insight from mean field theory.
\newblock {\em Can. J. Phys.}, {\bf 79}:1339--1359, (2001).

\bibitem{Review-Susceptibility-RE2Sn2O7}
V.~Bondah-Jagalu and S.~T. Bramwell.
%\newblock magnetic susceptinility study of the heavy rare earth stanate
%  pyrochlores.
\newblock {\em Can. J. Phys.}, {\bf 79}:1381--1385, (2001).

\bibitem{review-experimental-Harris-Zinkin}
M.~J. Harris and M.~P Zinkin.
%\newblock Frustation in the pyrochlore afs.
\newblock {\em Mod. Phys. Lett.}, {\bf 10}:417--438, (1996).

\bibitem{review-Magnetization-RE2Ti2O7}
S.~T. Bramwell, M.~N. Field, M.~J. Harris, and I.~P. Paskin.
%\newblock Bulk magnetization of the heavy rare earth titanate pyrochlores -- a
%  series of model frustrated magnets.
\newblock {\em J. Phys. Condens. Matter}, {\bf 12}:483--495, (2000).

\bibitem{review-spin-ice}
S.~T. Bramwell and M.~J.~P. Gingras.
%\newblock Spin ice state in frustrated magnetic pyrochlore materials.
\newblock {\em Science}, {\bf 294}:1495, (2001).

\bibitem{review-frus-magn}
C.~Lhuillier and G.~Misguich.
\newblock {\em Frustrated quantum magnets}.
\newblock {\em lecture notes of the Cargese summer school on { \it Trends in
  high magnetic field science} (may (2001))}, {\bf }.

\bibitem{Anderson-Spinel-1956}
P.~W. Anderson.
%\newblock {Ordering and AF in ferrites}.
\newblock {\em Phys. Rev.}, {\bf 102}:1008--1013, (1956).

\bibitem{Pauling}
L.~Pauling.
\newblock {\em The Nature of the chemical bond}.
\newblock Cornell University Press, Ithaca, 1938.

\bibitem{Baxter-Exact}
R.~J. Baxter.
\newblock {\em Exactly Solved Models in Statistical Mechanics}.
\newblock Academic Press, New--York, (1982).

\bibitem{CAF-n-comp-vec}
J.~N. Reimers, A.~J. Berlinski, and A.-C. Shi.
%\newblock Mfa to magnetic ordering in highly frustrated pyrochlores.
\newblock {\em Phys. Rev. B}, {\bf 43}:865--878, (1991).

\bibitem{CHAF-pyr-Reimers}
J.~N. Reimers.
%\newblock Absence of long-range order in a three-dimensional geometrically
%  frustrated antiferromagnet.
\newblock {\em Phys. Rev. B}, {\bf 45 }:45, (1992).

\bibitem{SL-Pyr-CAF}
B.~Canals and D.~Garanin.
%\newblock Spin liquid phase in the pyrochlore antiferromagnet: infinite
%  component spin.
\newblock {\em Can. J. Phys.}, {\bf 79}:1323--1331, (2001).

\bibitem{Class-geo-frus-AF-low-T}
R.~Moessner and J.~T Chalker.
%\newblock Low--temperature properties of classical geometrically frustrated
%  afs.
\newblock {\em Phys. Rev. B}, {\bf 58}:12049--12081, (1998).

\bibitem{CHAF-pyr-SL}
R.~Moessner and J.~T. Chalker.
%\newblock Properties of a classical spin liquid: the heisenberg pyrochlore af.
\newblock {\em Phys. Rev. Lett.}, {\bf 80}:2929, (1998).

\bibitem{MFQ-Tetr-Pyr}
A.J. Garcia-Adeva and D.L. Huber.
%\newblock Quantum tetrahedral mf theory of the magentic susceptibility for the
%  pyrochlore lattice.
\newblock {\em Phys. Rev. Lett.}, {\bf 85}:4598, (2000).

\bibitem{QSL-HAF-Pyr}
B.~Canals and C.~Lacroix.
%\newblock Quantum spin liquid: Study of the quantum heisenberg af on the three
%  dimensional pyrochlore lattice.
\newblock {\em Phys. Rev. B}, {\bf 61}:1149, (2000).

\bibitem{QAF-3D-Pyr}
B.~Canals and C.~Lacroix.
%\newblock Pyrochlore af: a 3d quantum spin liquid.
\newblock {\em Phys. Rev. Lett.}, {\bf 80}:2933, (1998).

\bibitem{AFQ-pyr}
H.Tsunetsugu.
%\newblock Af quantum spins on a pyrochore lattice.
\newblock {\em J. Phys. Soc. Jpn.}, {\bf 70}:640--643, (2001).

\bibitem{Tetrahedra-QHAF}
M.~Elhajal, B.~Canals, and C.~Lacroix.
%\newblock Comparison of several tetrahedra--based lattices.
\newblock {\em Can. J. Phys.}, {\bf 79}:1353--1357, (2001).

\bibitem{Pyrochlore-Plaquette-Isoda-Mori}
M.~Isoda and S.~Mori.
%\newblock Valence-bond crystal and anisotropic excitation spectrum on 3-
%  dimensionally frustrated pyrochlore.
\newblock {\em J. Phys. Soc. Jp.}, {\bf 67}:4022, (1998).

\bibitem{Harris-Berlinsky-Bruder-1991}
A.~B. Harris, A.~J. Berlinsky, and C.~Bruder.
\newblock {\em J. Appl. Phys.}, {\bf 96}:5200, (1991).

\bibitem{Pyrochlore-Singlet-Exc}
E.~Berg, E.~Altman, and A.~Auerbach.
%\newblock Singlet excitations in pyrochlore: A study of quantum frustration.
\newblock {\em Preprint}, {\bf }.

\bibitem{Classical-generalized-constant-coupling}
A.J. Garcia-Adeva and D.~L. Huber.
\newblock {\em Phys. Rev. B}, {\bf 63}:140404, (2001).

\bibitem{Effective-Renormalization-Group}
A.J. Garcia-Adeva and D.~L. Huber.
%\newblock Critical behaviour of 2 and 3 d f and af spin-ice systems in the
%  framework of the effective field renormalization group.
\newblock {\em Phys. Rev. B}, {\bf 64}:14418, (2001).


\bibitem{Spin-Ice-Ho2Ti2O7}
M.~J. Harris, S.~T. Bramwell, D.~F. McMorrow, T.~Zeiske, and K.~W. Godfrey.
%\newblock Geometrical frustration in the ferromagnetic pyrochlore ho2ti2o7.
\newblock {\em Phys. Rev. Lett.}, {\bf 79}:2554, (1997).

\bibitem{Ziman-CVA}
J.~M. Ziman.
%\newblock {\em {\it Models of Disorder}}.
\newblock Cambridge University Press, Cambridge, 1979.

\bibitem{CVA-Ising}
E.N.M. Cirillo, G.~Gonnella, M.~Troccoli, and A.~Maritan.
%\newblock Correlation functions by cluster variation method for ising model
%  with nn, nnn and plaquette interactions.
\newblock {\em Journ. Stat. Phys.}, {\bf 94 }:67--89, (1999).

\bibitem{CVM-spin-ice}
S.~Yoshida, K.~Nemoto, and K.~Wada.
%\newblock Application of the cvm to spin ice systems on the pyrochlore pattice.
\newblock {\em J. Phys. Soc. Jap.}, {\bf 71 }:948--954, (2002).

\bibitem{Husimi-SL}
P.~Chandra and B.~Doucot.
%\newblock Spin liquids on the husimi cactus.
\newblock {\em J. Phys. A: Math. Gen.}, {\bf 27}:1541--1556, (1994).

\bibitem{Cayley-tree-FI-Glass}
R.~Melin, J.~C.~Angles d'Auriac, P.~Chandra, and B.~Doucot.
%\newblock Glassy behaviour in the ferromagentic ising model on a cayley tree.
\newblock {\em J. Phys. A: Math. Gen.}, {\bf 29 }:5773--578, (1996).

\bibitem{Henning}
J.~C.~M. Henning.
%\newblock direct determination of weak excahnge interactionbs in cr3+--doped
%  spinel znga2o4.
\newblock {\em Phys. Rev. B}, {\bf 21}:4983--4995, (1980).

\bibitem{Single-Ion-Anis-Pyr}
R.~Moessner.
%\newblock Relief and generation of frustration in pyrochlore magnets by
%  single--ion anisotropy.
\newblock {\em Phys. Rev. B}, {\bf 57}:R5587--R5589, (1998).

\bibitem{Luciano-Peixoto}L.~Peixoto, private communication.

\bibitem{1D-AfH-s=1/2-alternating}
D.~C. Johnson, R.~K. Kremer, M.~troyer, X.~Wang, A.~Klumper, S.~L. Dud'ko,
  A.~F. Panchula, and P.~C. Canfield.
%\newblock Thermodynamics of s=1/2 af uniform and alternating heisenberg chain.
\newblock {\em Phys. Rev. B}, {\bf 61 }:9558, (2000).

\bibitem{CAF-HTE-KAgome}
A.~B. Harris, C.~Kallin, and A.~J. Berlinsky.
%\newblock Possoble neel orderings of the kagome af.
\newblock {\em Phys. Rev. B}, {\bf 45}:2899--2919, (1992).

\bibitem{HTE-ZnxCd1-xCr2Se4}
N.~Bezakour, M.~Hamedoun, M.~Houssa, A.~Hourmatallah, and F.~Mahjoubi.
%\newblock Study of critical properties of b-spinel zncdcr2se4.
\newblock {\em M. J. Condensed Matter}, {\bf 1}:2, (1999).

\bibitem{Crit-temp-interpolated-Rushbrooke-Wood}
G.~S. Rushbrooke and P.~J. Wood.
\newblock {\em Mol. Phys.}, {\bf 1}:257, (1958).

\bibitem{CHAF-Pyr-Entropy-Energy}
C.~Pinettes, B.~Canals, and C.~Lacroix.
%\newblock Classical heisenberg antiferromagnet away from the pyrochlore lattice
%  limit: entropic versus energetic selection.
\newblock {\em Phys. Rev. B}, {\bf 66 }:24422, (2002).

\bibitem{Frus-HAF-Pyr}
A.~Koga and N.~Kawakami.
%\newblock Frustrated heisenberg af on a pyrochlore lattice.
\newblock {\em Phys. Rev. B}, {\bf 63 }:144432, (2001).

\bibitem{HTE-ZnCdCr2Se4-2}
N.~Benzakour, M.~Hamedoun, M.~Houssa, A.~Hourmatallah, and F.~Mahjoubi.
%\newblock Crit. prop. of the magn. disordered b-spinel zncdcr2se4.
\newblock {\em phys. stat. sol. b}, {\bf 212 }:335--342, (1999).

\bibitem{CuGe2O4-spinel-1D}
T.~Yamada, Z.~Hiroi, M.~Takano, M.~Nohara, and Hidenori Takagi.
%\newblock Spin-1/2 heisenberg antiferromagnetic chains in the frustrated spinel
%  lattice of gecu2o4.
\newblock {\em J. Phys. Soc. Jpn.}, {\bf 69}:1477--1483, (2000).

\bibitem{LiV2O4-Pyrochlore}
V.~Eyert, K.-H. Hoeck, S.~Horn, A.~Loidl, and P.~S. Riseborough.
%\newblock Electronic structure and magnetic interactions in liv2o4.
\newblock {\em Europhys. Lett.}, {\bf 46}:762--767, (1999).


\bibitem{GGG-Petrenko-1}
O.~A. Petrenko, C.~Ritter, M.~Yethiraj, and D.~Paul.
%\newblock Spin-liquid behavior of the gadolinium gallium garnet.
\newblock {\em Physica B}, {\bf 241-243 }:727--729, (1997).

\bibitem{GGG-Petrenko-2}
O.~A. Petrenko, D.~M. Paul, C.~Ritter, T.~Zeiske, and M~Yethiraj.
%\newblock Magnetic frustration and order in gadolinium gallium garnet.
\newblock {\em Physica B}, {\bf 266}:41, (1999).

\bibitem{Field-induced-order}
M.E. Zhitomirsky, A.~Honecker, and O.A. Petrenko.
%\newblock Field induced ordering in highly frustrated afs.
\newblock {\em Phys. Rev. Lett.}, {\bf 85}:3269--3273, (2000).

\bibitem{Gd2Ti2O7-Mag-Field-Ind-Phase-Tran}
A.~P Ramirez, B.~S. Shastry, A.~Hayashi, J.~J. Krajewski, D.~A. Huse, and R.~J.
  Cava.
%\newblock multiple field induced phase transition in a
%  geometrically--frustrated dipolar magnet -- gd2ti2o7.
\newblock {\em Phys. Rev. Lett.}, {\bf 89}:67202, (2002).

\end{thebibliography}

\newpage
\section*{}
\begin{centering}
\begin{table}[t]
\caption[table 1]{The energy levels of the spins on a  tetragonally distorted
tetrahedron for $s=1/2$.
\label{table:-energy-levels-s=1/2}}
\center{
\begin{tabular}{|l|c||l|c|}
\hline
S M L R & E(S,M,L,R)                &S M L R      & E(S,M,L,R) \\
\hline
0  0 0 0      & $ 0 $               &2 $\pm$1 1 1 & $6J-3D + 4\Delta$    \\
1  0 1 0      & $2(J+D) + 2\Delta$  &2 0 1 1      & $6(J+D) + 4\Delta$   \\
1  0 0 1      & $2(J+D) + 2\Delta$  &0 0 1 1      & $4\Delta $           \\
1 $\pm$1 1 0  & $2J - D + 2\Delta$  &1 $\pm$1 1 1 & $J-D + 4\Delta $    \\
1 $\pm$1 0 1  & $2J - D + 2\Delta$  &1 0 1 1      & $2(J+D) + 4\Delta$  \\
2 $\pm$2 1 1  & $6J - 6D + 4\Delta$ &             &  \\
\hline
\end{tabular}
}
\end{table}
\end{centering}

\newpage
\section*{}

\begin{centering}
\begin{table}[t]
\caption[table 2]{   Energies   in   units of $J$ obtained from
approximate  calculations  on a quintet cluster of tetrahedra  for
AF exchange    interactions.  For details   see   the  text.   The
degeneracy and spin  of each  state are  indicated in parenthesis.
In the row $NS$  we indicate  the number of total spin states  and
the number of local spin states  present in each calculation.  The
total  number of  states with $M_Q = 0$ is  12870.   The  energies
should be corrected by -12. The exact ground state energy has
been calculated by~\cite{Luciano-Peixoto}.
\label{table:-energy-levels-quintet-cluster}
}
\center{
\begin{tabular}{|c|ccccc|c|}
\hline
$(S,D,Q,M_Q)$&(0,0,0,0)  &(1,1,1,0)    &(1,1,1,0)    &(1,1,1,0)
&(1,1,1,0)&exact\\
$S_{sum}$  &0            &1            &2            &3            &4&\\
 $NS = $   &16/1296      &112/9232     &520/9232     &1192/9232
&1597/9232&12870\\
\hline
1
&-0.188(16;0)&-0.235(16;0)&-0.830(16;0)&-0.830(16;0)&-0.856(16;0)&-0.82(16;0)\\
2          & --
&1.500(48;1)&0.939(48;1)&0.767(32;1)&0.712(28,1)&0.69(48;1)\\
3	   &  --
&1.812(32;1)&1.690(32;1)&0.938(16;1)&0.750(1,1)&1.23(32;1))\\
4	   &  --
&2.797
(16;1)&2.000(32;0)&1.448(32;1)&0.755(1,1)&1.77(32;0)\\
\hline
$E_{ST}$   &  --         &1.733      &1.768       &1.597
&1.568&1.51\\
\hline
\end{tabular}
}
\end{table}
\end{centering}
\newpage
\begin{centering}
\begin{table}[t]
\caption[table 3]{The ground state energy per site and singlet triplet
gap $\Delta E_{ST}$, both in units of 
$J$. 
\label{table:-groundstate-STgap}
}
\vspace{.2 mm}
\center{
\begin{tabular}{|cc|c|c|}
\hline
$E_G$& $\Delta E_{ST}$&method&ref. \\ 
\hline
-0.96        & --     &$J^{\prime}/J$ Expansion&\cite{Frus-HAF-Pyr}\\
%A. Koga and N. Kawakami
-0.916       & 1.688  &$J^{\prime}/J$ Expansion&\cite{Pyrochlore-Plaquette-Isoda-Mori}\\
%Isoda-Mori
-1.00        &  --    &$J^{\prime}/J$ Expansion&\cite{Tetrahedra-QHAF}\\
%Elhajal et al 
-1.144       &  --    &Spin-Waves              &\cite{Harm-SW-QAF-pyr}\\
%Sobral
-0.975       & --     &Mean Field              &\cite{Harris-Berlinsky-Bruder-1991}\\
-1.0?        &   1.10 &$J^{\prime}/J$ Expansion&\cite{AFQ-pyr}\\
%Tsunetsugu
-1.12        & --     &$J^{\prime}/J$ Expansion&\cite{QSL-HAF-Pyr}\\
%Canals and C. Lacroix PRB
-1.10        & 1.4    &Exact 16 site           &\cite{QAF-3D-Pyr}\\
%Canals and C. Lacroix PRL
-1.12?       & 0.8    &SuperTetrahedron: 16 sites&\cite{Pyrochlore-Singlet-Exc}\\
%Berg et al
\hline
\end{tabular}
}
\end{table}
\end{centering}

\newpage
\section*{}

\begin{centering}
\begin{table}[t]
\caption[table 4]{The b coefficients 
for $s=1/2$ for the tetrahedron as basis cluster.}
\label{table:b-coeff}
\center{
\begin{tabular}{|c||c |ccc |ccc |c |ccc |ccccc|}
\hline
$S$ &   0& 1&  1& 1&  1&  1&  1& 0&  1& 1& 1&  2&  2& 2& 2& 2\\
$M$ &   0& -1& 0& 1& -1&  0&  1& 0& -1& 0& 1& -2& -1& 0& 1& 2\\
$L$ &   0& 0&  0& 0&  1&  1&  1& 1&  1& 1& 1&  1&  1& 1& 1& 1\\
$R$ &   0& 1&  1& 1&  0&  0&  0& 1&  1& 1& 1&  1&  1& 1& 1& 1\\
\hline
$m=-1/2$& 1& 0& 0& 0& 1/2&  0&-1/2&  0&	1/4& 0&-1/4& 1/2& 1/4&  0&-1/4&-1/2  \\
$m=1/2$ & 1& 0&	0& 0&-1/2&  0& 1/2&  0&-1/4& 0& 1/4&-1/2&-1/4&	0& 1/4&	1/2  \\
\hline
\end{tabular}
}
\end{table}
\end{centering}

\newpage
\section*{}

\begin{centering}
\begin{table}[t]
\caption[table 5]{Critical coordination number $Z_{crit}$ and critical
temperature   in  the  pair--CVA   $T_{c,pair}/T_{c,MF}$   the tetrahedron--CVA
$T_{c,H}/T_{c,MF}$ for ferromagnetic {\it  nn} Heisenberg exchange $J$  and the
ratio of   the Ising and Heisenberg  critical temperature $T_{c,I}/T_{c,H}$  on
the pyrochlore lattice -- the Curie--Weiss  mean field value is: $k_B T_c = 2
J Z  s(s+1)/3$ for  $Z=6$.  The critical temperature in the Heisenberg  limit
(zero--field  zero) ($k_B T_{c,H}/J$)  and in the Ising limit ($k_B T_{c,I}/J$) for
$D_s/J = 5000$.  The ratio  $T_{c,I}/T_{c,H}$    is  in    the     mean   field
approximation this ratio is $(1+1/S)/3$.  The high temperature  expansion  is
derived from the formula of Rushbrooke and Wood
\cite{Crit-temp-interpolated-Rushbrooke-Wood}. 
In the last  two rows   we give  the  characteristic  critical   values  for the
tetrahedral zero--field $D_T$.
\label{table:-critical-temperature-tetrahedra} 
}
\center{
\begin{tabular}{|l|cccc ccc|}
\hline
$ s=$           & 1/2   & 1      & 3/2   & 2     & 5/2   & 3     & 7/2 \\
\hline
\multicolumn{8}{l}{Pair--CVA with pairs, ferromagnetic Heisenberg spins:}\\
\hline
$Z_{crit}$      &  4    & 3      & 2.66  & 2.4   &2.3    &2.26   & 2.25\\
$T_{c,pair}/T_{c,MF}$&0.607  & 0.737  & 0.773 & 0.788 & 0.796 & 0.801 & 0.804 \\
\hline
\multicolumn{8}{l}{Tetrahedron--CVA with ferromagnetic Heisenberg (H) and
Ising (I) spins:}\\
\hline
$k_B T_{c,H}/J$  & 0     & -4.664 & -9.636&-16.001&-23.980&-32.955&-43.566  \\
$k_B T_{c,I}/J$  & 0     & -8.808 &-19.818&-35.234&-55.054&-79.277&-107.906 \\
$T_{c,H}/T_{c,MF}$& 0     & 0.583  & 0.642 & 0.667 & 0.679 & 0.687 & 0.692 \\
$T_{c,I}/T_{c,H}$ &  NA   & 1.888  & 2.056 & 2.208 & 2.295 & 2.406 & 2.477 \\
\hline
\multicolumn{8}{l}{High temperature expansion Heisenberg model $Z=6$
ferromagnetic:}\\
\hline
$k_B T_c/J$     & 0.629 & 0.684  & 0.699 & 0.705 & 0.709 & 0.711 & 0.712   \\
\hline
\end{tabular}
}
\end{table}
\end{centering}

\newpage
\section*{}

\begin{centering}
\begin{table}[t]
\caption[table:6]{
In this table we   give the characteristic critical values for the
tetrahedral   zero--field    $D_{T,max}$  and the maximum critical
temperature $T_{max}$ which characterizes  the re--entrant regiion
for   $s=1/2$             to     7/2.         See           figure
\ref{figure:-crit-temp-single-ion-anis-s=3/2}.
}
\label{table:-reentrant}
\center{
\begin{tabular}{|l|cccc ccc|}
\hline 
$T_{max}/J$(max) &NA & -2.647   & -5.128&  -8.143&-12.400&-15.662    & -20.137    \\
$D_{T,max}/J\times  100$(max)&  NA&   -3.695& -6.312&  -7.985&   -9.378&
-9.982 & -10.465\\ 
\hline
\end{tabular}
}
\end{table}
\end{centering}

\pagebreak

\section*{}

\begin{figure}

\caption[fig:01]{\label{figure:-pyrochlore-structure}          The
pyrochlo (=B sublattice of  the spinel structure) lattice. We have
indicated the  Corner--Sharing--Tetrahedra in blue, and some
of the hexagon loops. The cubes (black thin lines)
are drawn to guide the eye}

\caption[fig:02]{\label{figure:-tetrahedron}The   tetrahedron with
site   and bond indices as used   in the  main text.  The    local
trigonal $\langle\!  111\!   \rangle$  quantization  axes and  a
tetragonal axis ($C_2$ or $S_4$) are also indicated.}

\caption[fig:03]{\label{figure:-energy-levels-s=1/2}In        this
figure   we illustrate   the energy levels of  spin  states   on a
tetrahedron  for $s=1/2$. The exchange  interaction is  $J=1$, the
reduced tetragonal   anisotropy is $d_T = D_T/J$,  and the reduced
exchange anisotropy is $\delta = \Delta/J$.  The  reduced magnetic
field is $h = \mu_B   H/J$.  The  magnetic    field is along   a
tetragonal axis. The ($L,R$)  character of the states is indicated
in parenthesis, the degeneracy  is  indicated in brackets. For the
energies see also table: \ref{table:-energy-levels-s=1/2}.}

\caption[fig:04]{\label{figure:-phase-diagrams}  The ground  state
phase diagram for ferromagnetic (F) and  anti ferromagnetic   (AF)
{\it  nn}  exchange     interactions      for   $s=1/2$,    in the
$d_T$--$\delta$ plane.   The numbers in parenthesis  indicate  the
$L,R$ character of the  ground state. The equations separating the
various ground states are given in the main text.  }

\caption[fig:05]{
\label{figure:-critical-temperature-ferromagnetic-order-coordination-number}
The reduced critical  temperature    $k_B  T_c/J$ as a function of
$C_t = 2(Z_t-1)/Z_t$ for ferromagnetic   nn exchange  for $s=1/2$;
$Z_t$ is the  coordination number. The case of 2, 3  and  4 CST is
$C_t =1$, 4/3 and 1.5, respectively, are indicated as squares.}

\caption[fig:06]{
\label{figure:-reentrant-region-ferromagnetic-nn-interactions}
Detail   of  the ferromagnetic  phase  diagram for $s=1/2$  in the
$d_T$--$\delta$ plane. We have indicate in black  the region  with
re--entrant  behaviour,  which separates  the singlet  (S)  ground
state region from the ferromagnetic (F) ground state region.}

\caption[fig:07]{
\label{figure:-critical-temperature-tetrahedron-anisotropy}    The
reduced   critical  temperature  $k_B T_c/J$ as a function  of the
reduced tetrahedral anisotropy   $d_T  = D_T/J$ for various values
of the reduced exchange anisotropy  $\delta = \Delta/J   = 0., -1,
-1.5,   -2.0,  -2.5,  -3.0$  for $s=1/2$.      Paramagnetic   (P),
Ferromagnetic (F) and   singlet--like  or  collective paramagnetic
(S) behaviour   of the spin system is indicated.  See also  figure
\ref{figure:-reentrant-region-ferromagnetic-nn-interactions}   and
table~\ref{table:-critical-temperature-tetrahedra}. In  this graph
we have also indicated the  curves, for  the same  set of $\delta$
values, below which the entropy becomes negative: $T_S$.}

\caption[fig:08]{\label{figure:-entropy-ferro-s=1/2}The  number  of
states   accessible   as a  function  of the temperature       for
ferromagnetic  exchange    (F)  and anti--ferromagnetic   exchange
(AF).}

\caption[fig:09]{\label{figure:-entropy-magnetic-field-s=1/2}The
temperature   $T_S$ for   which   the  entropy     vanishes    for
non--magnetic  spin system for ferromagnetic exchange  $J = -1$ as
a function    of the   tetragonal anisotropy  $d_T$, for  magnetic
fields  $h =0$,  $1$ and  $2$,  along the line   A-C in the  phase
diagram~\ref{figure:-phase-diagrams}. }
\end{figure}

\begin{figure}
\caption[fig:10]{
\label{figure:-crit-temp-single-ion-anis-s=3/2}
The reduced critical temperature $k_B  T_c/J$ as a function of the
reduced tetragonal   anisotropy   $d_T =  D_T/J$  for  $s=3/2$ and
reduced tetrahedral anisotropy.   Paramagnetic  (P), Ferromagnetic
(F) and singlet--like  or collective paramagnetic (S) behaviour of
the spin   system  is  indicated.    The maximum   temperature and
anisotropy                values               given            in
table~\ref{table:-critical-temperature-tetrahedra}.}           are
indicated.

\caption[fig:11]{ \label{fig:-ferro-M-H} The magnetization
$\cal{M}$ as a function of the magnetic field $h$ for
ferromagnetic exchange $J = -1$ using
equation~\ref{equation:-magnetization-s=1/2}, for three
temperatures $T=0.1$, $1$ and $5$, for various values of the
tetrahedral anisotropy $|d_T| = 0.05, 0.1, 0.25, 0.5, 1.0, 2.0$
for $\delta=-3$ (line A-C in fig.~\ref{figure:-phase-diagrams} )
and $\delta = 1$ (line B-D in fig.~\ref{figure:-phase-diagrams}).
A: $\delta = -3$, $d_T$ positive; B: $\delta = 1$, $d_T$:
negative; C: $\delta =-3$, $d_T$: negative; D:$\delta = 1$,
$d_T$: positive. For details see the text. }

\caption[fig:12]{\label{figure:-reentrant-M-T-s=1/2}The
magnetization      $\cal{M}$  in  the re--entrant  region      for
ferromagnetic  {\it nn} exchange $J=-1$ for $s=1/2$ in small field
$h = 0.001$  for  $\delta    = -3$   for  various  values  of  the
tetragonal  anisotropy from  $d_T=-0.85$ to $-1.075$ with steps  of
$-0.025$. Units used: $|J|$. }

\caption[fig:13]{   \label{fig:-M-T} The magnetization ${\cal{M}}$
as a function of  temperature  for ferromagnetic exchange $J = -1$
using   equation~\ref{equation:-magnetization-s=1/2},      for the
magnetic fields $h=H/|J| = 0.01, 1, 2,  3, 5, 10  $,  for  various
values  of the tetrahedral anisotropy $d_T = 0.05, 0.1, 0.25, 0.5,
1.0, 2.0$     for    $\delta      =   1$    (line      B-D      in
fig.~\ref{figure:-phase-diagrams}).     For details  see the text.
Units used: $|J|$.}

\caption[fig:14]{\label{fig:-specific-heat-reentrant}          The
specific  heat  as a function  of  temperature for small  magnetic
field   in the re--entrant     regiion      for $\delta = -3$   and
$d_T=0.925$.}

\end{figure}
\pagebreak
\section*{}
\pagebreak

\epsfverbosetrue
\epsfysize=6in
\epsffile{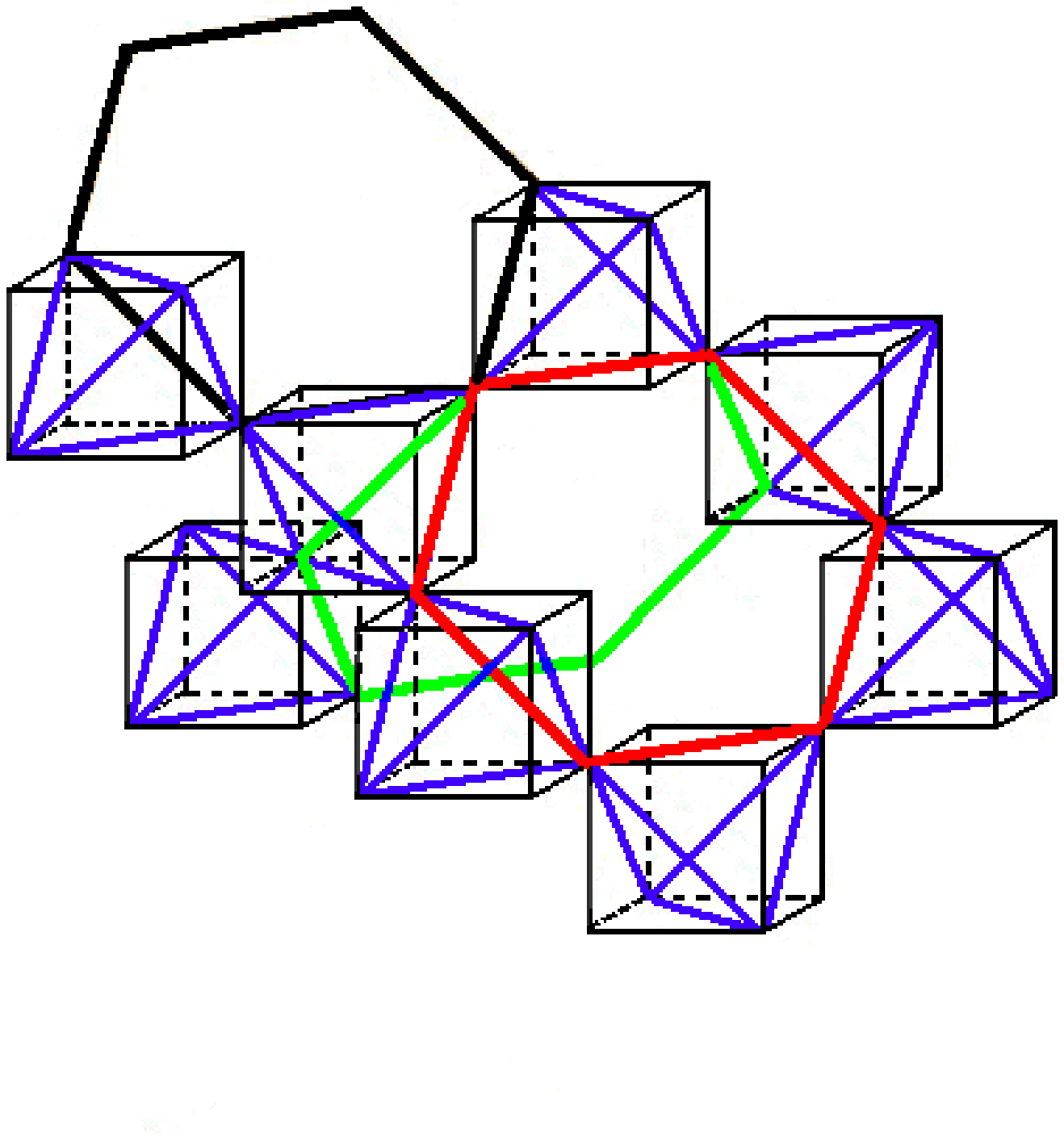}

\newpage

\epsfysize=5in
\epsffile{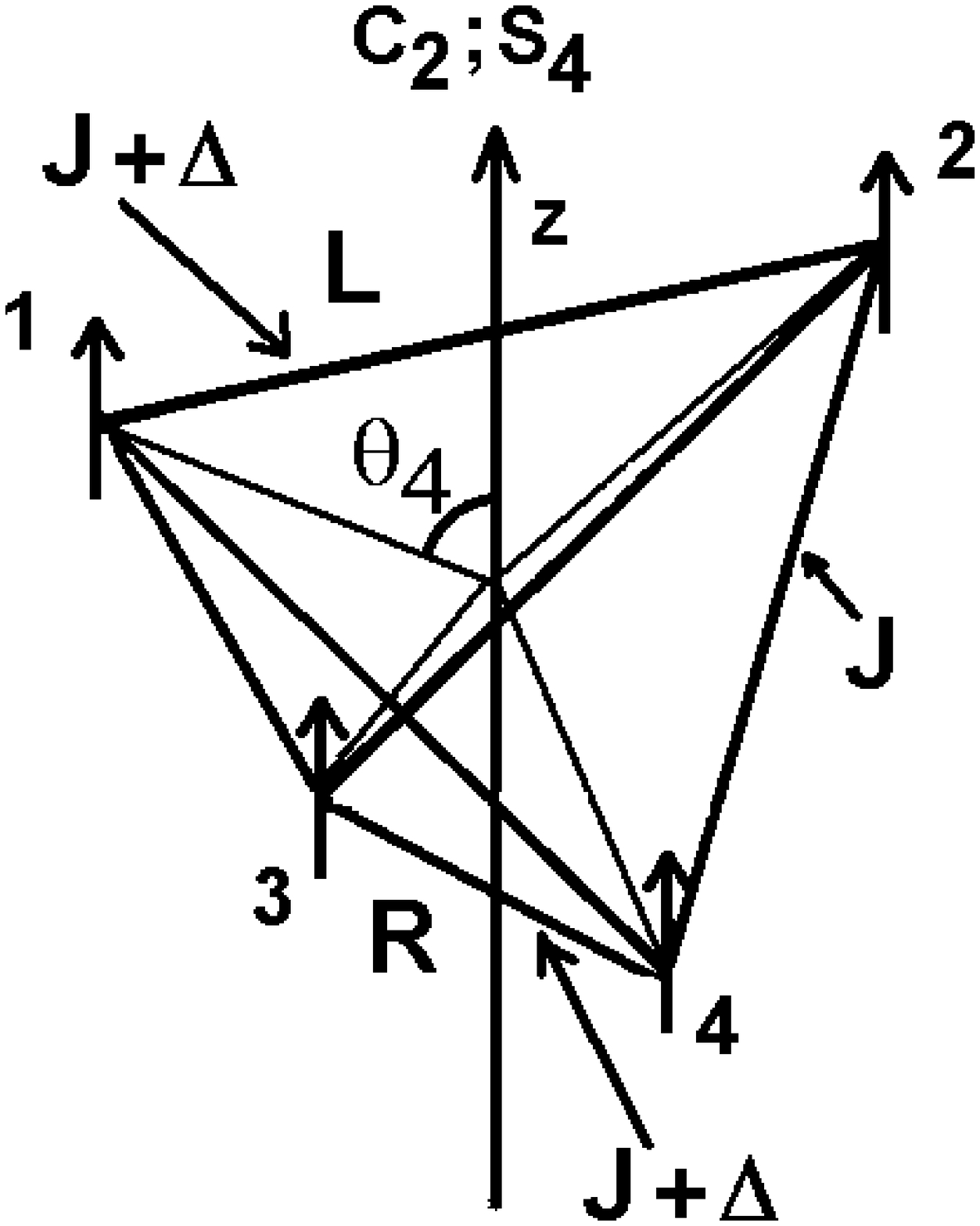}

\newpage

\epsfysize=7in
\epsffile{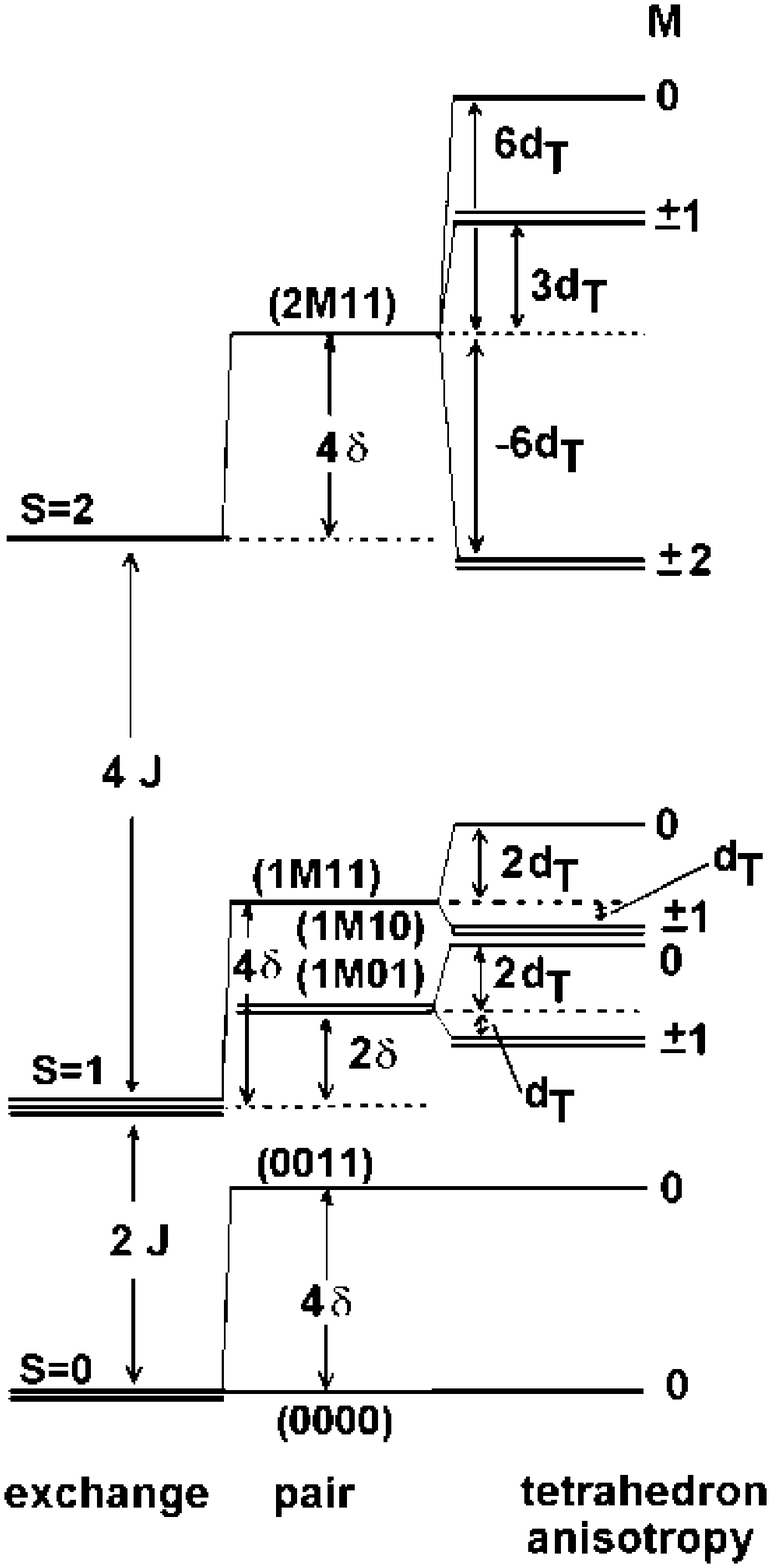}

\newpage

\epsfysize=5in
\epsffile{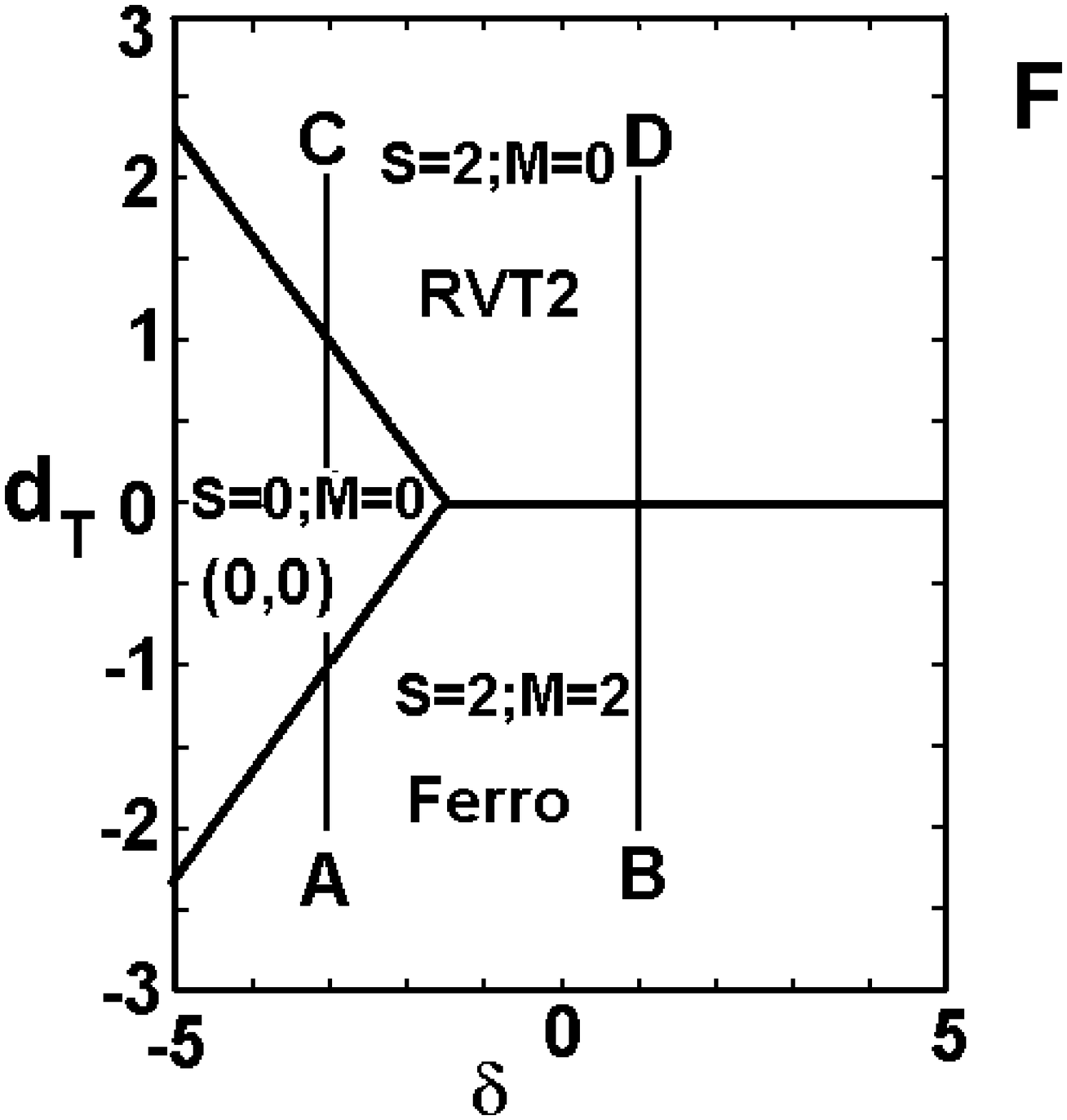}

\newpage

\epsfysize=5in
\epsffile{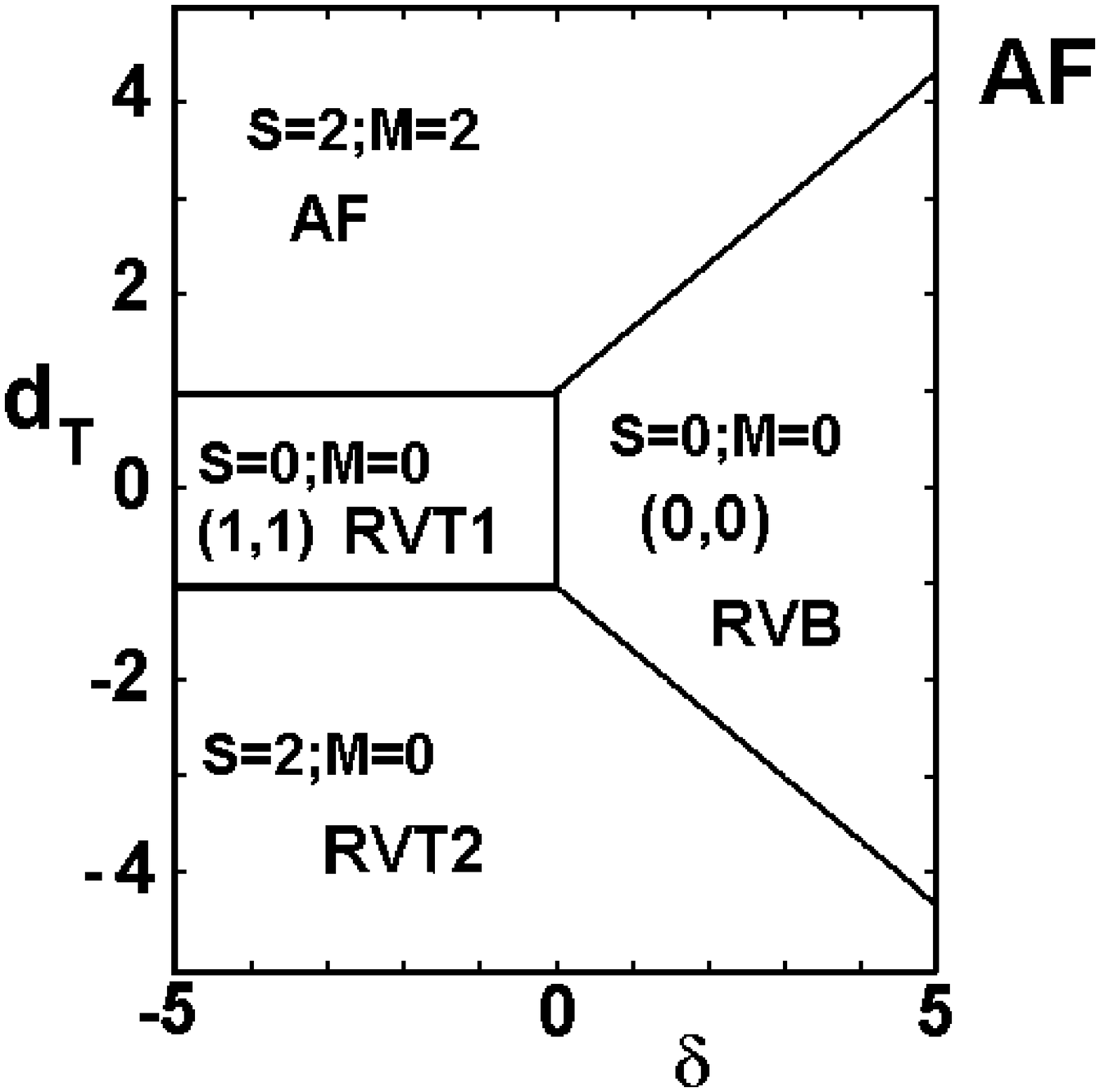}

\newpage

\epsfysize=4in
\epsffile{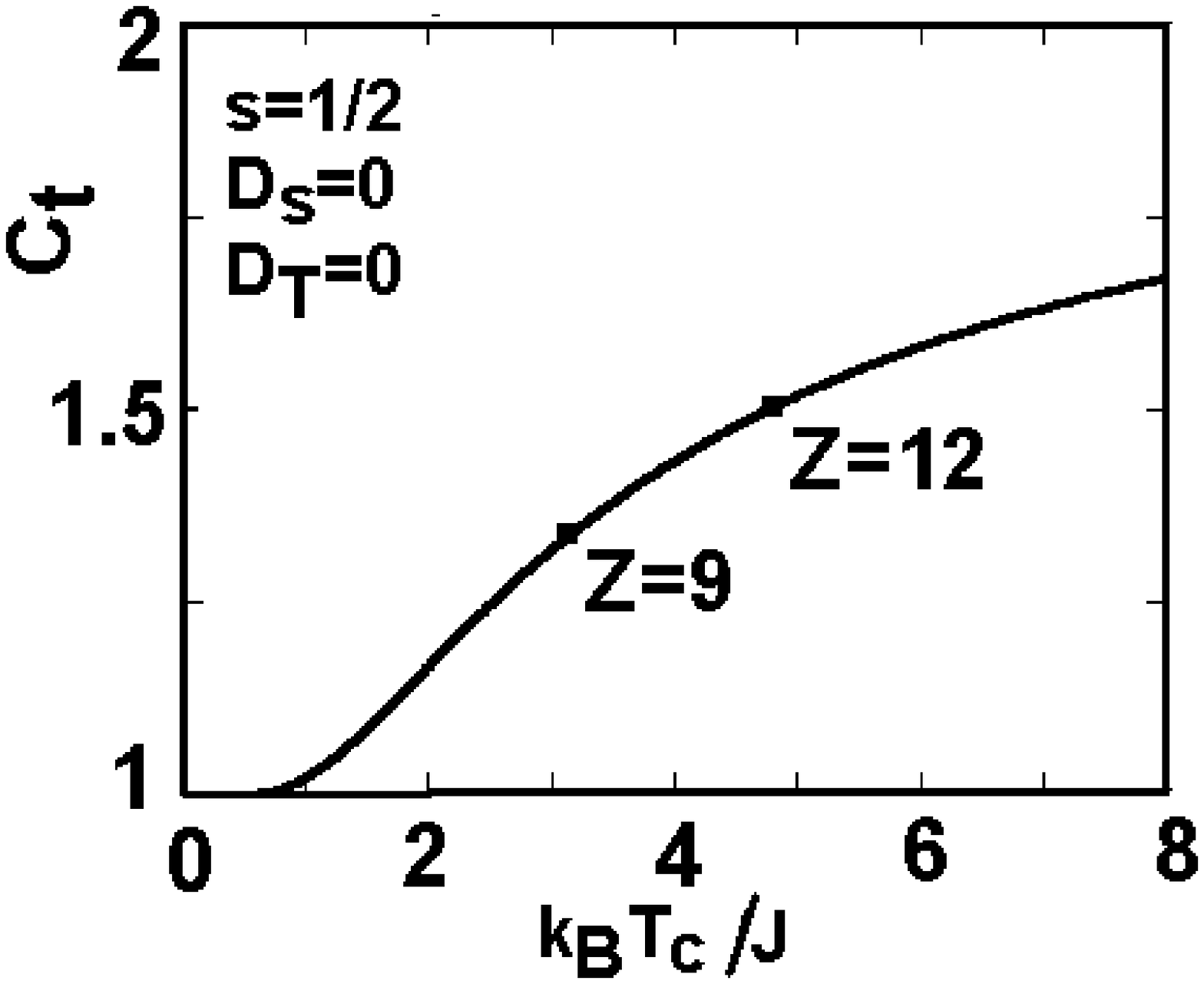}

\newpage

\epsfysize=4in
\epsffile{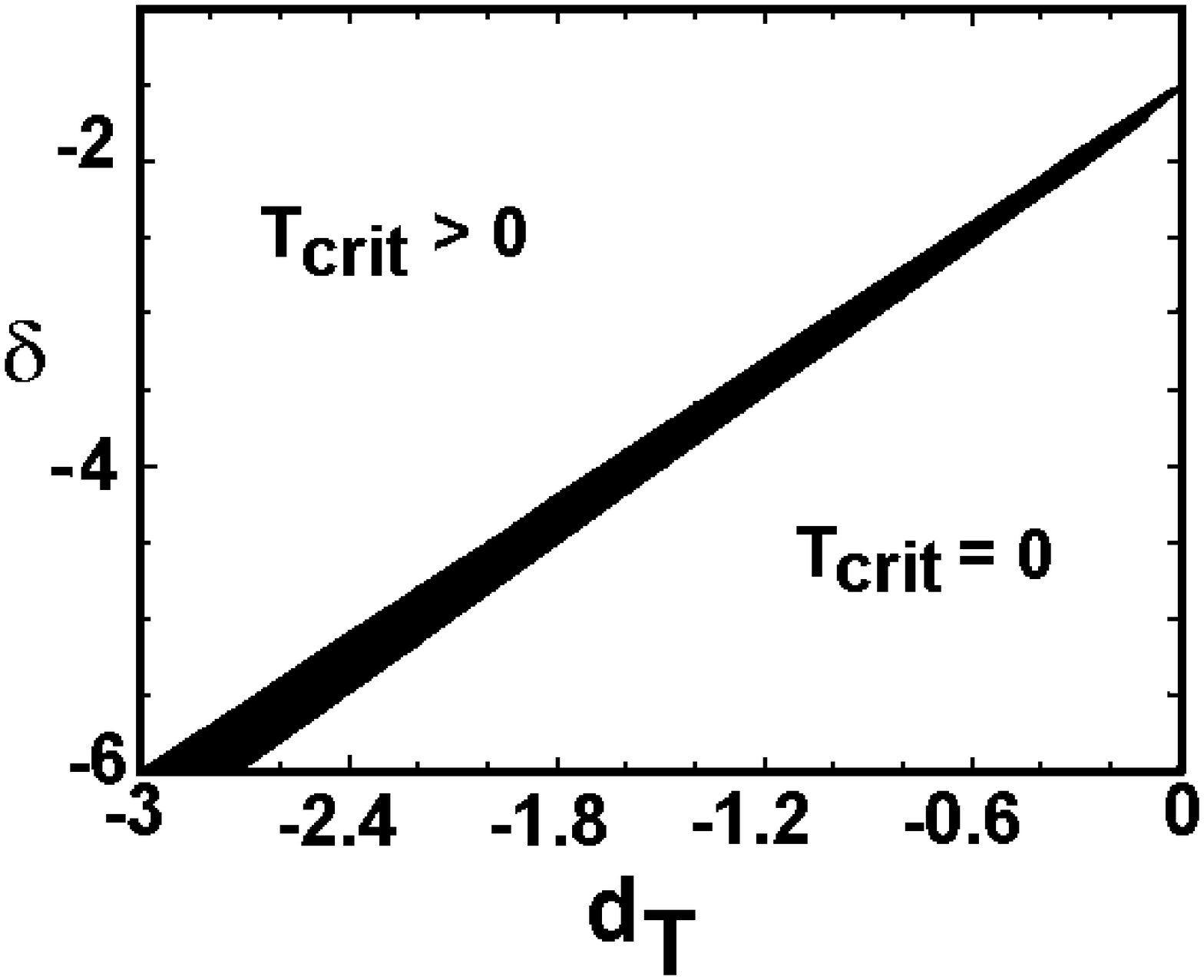}

\newpage

\epsfysize=5in
\epsffile{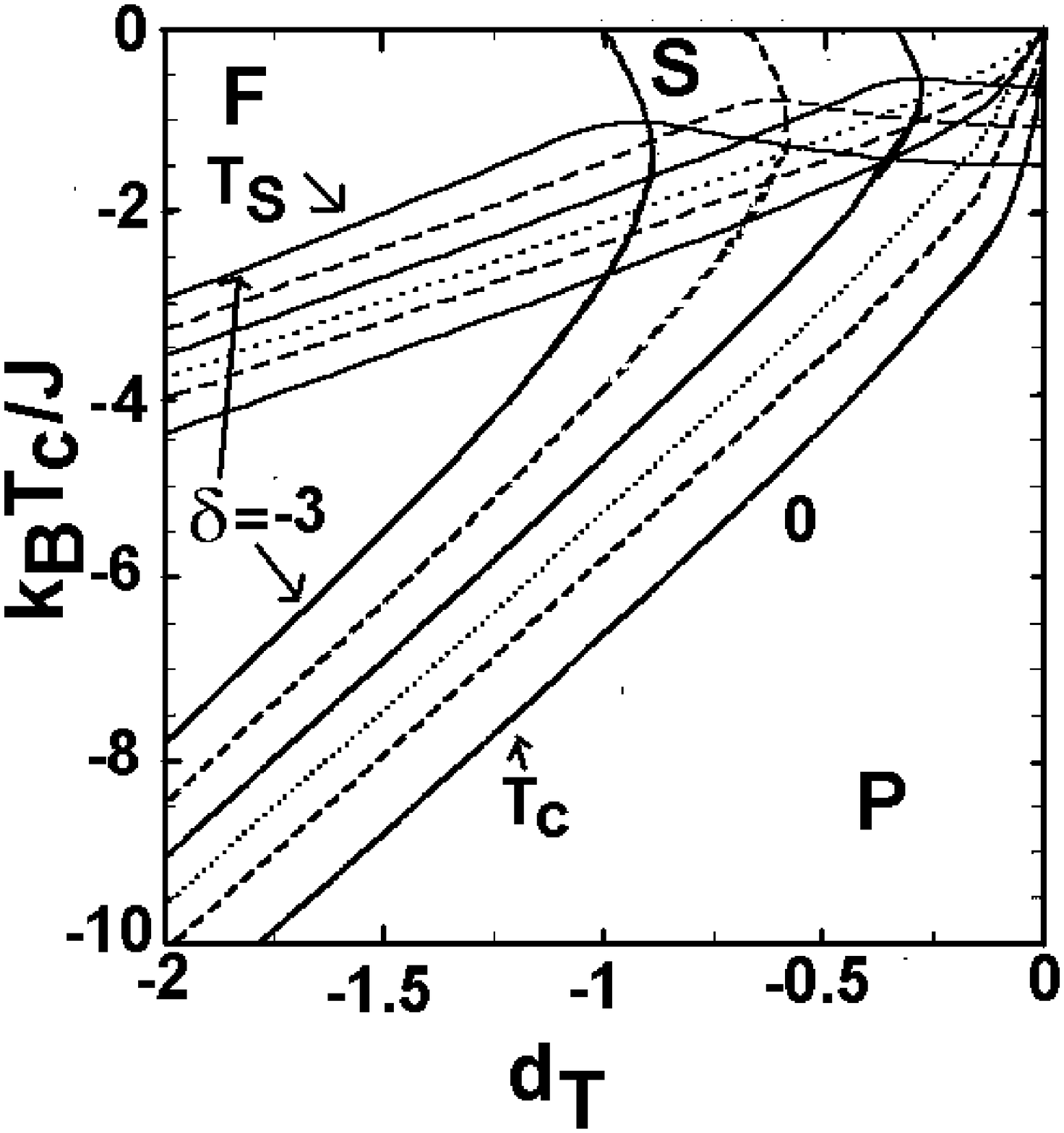}

\newpage

\epsfysize=5in
\epsffile{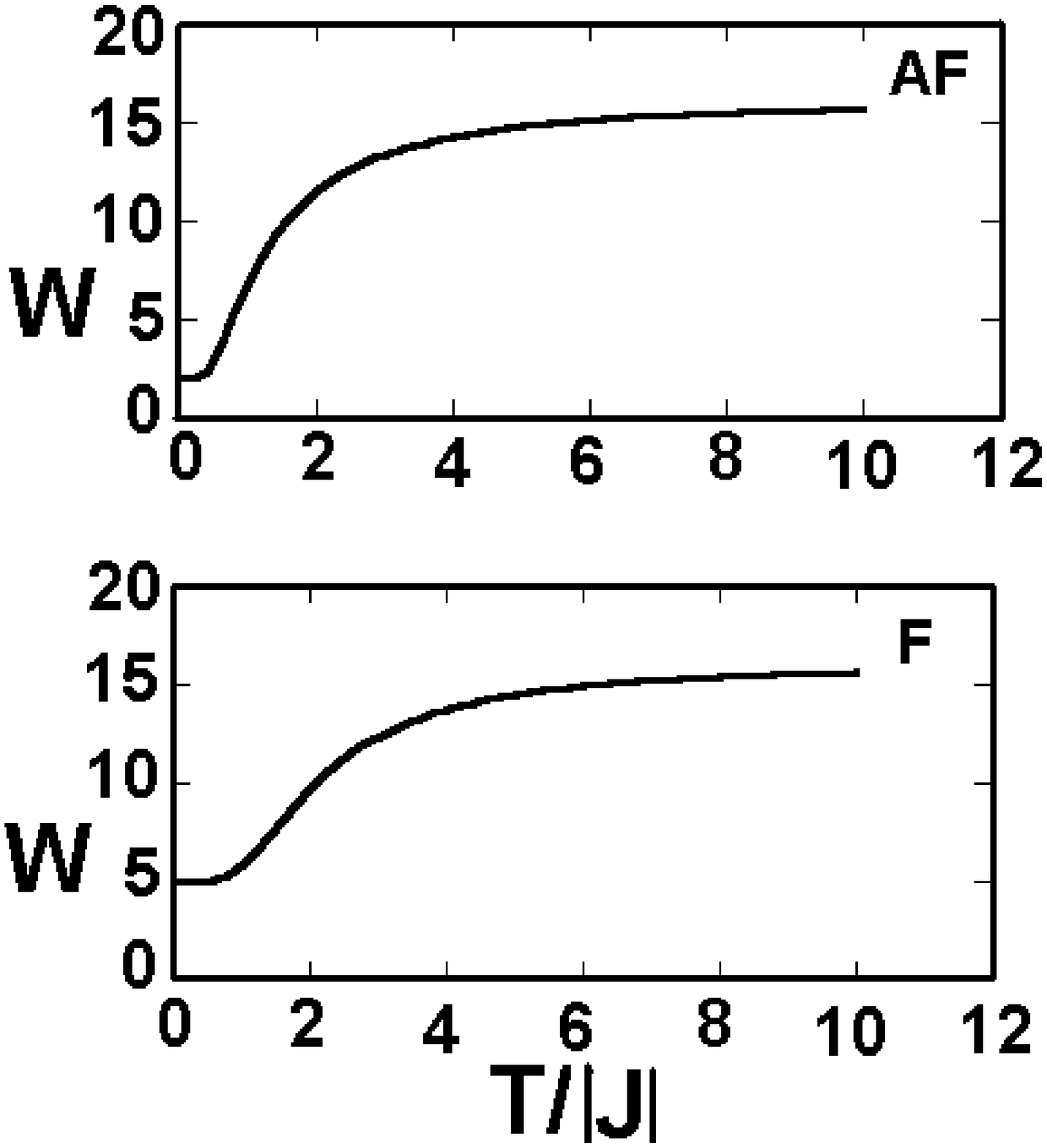}

\newpage

\epsfysize=5in
\epsffile{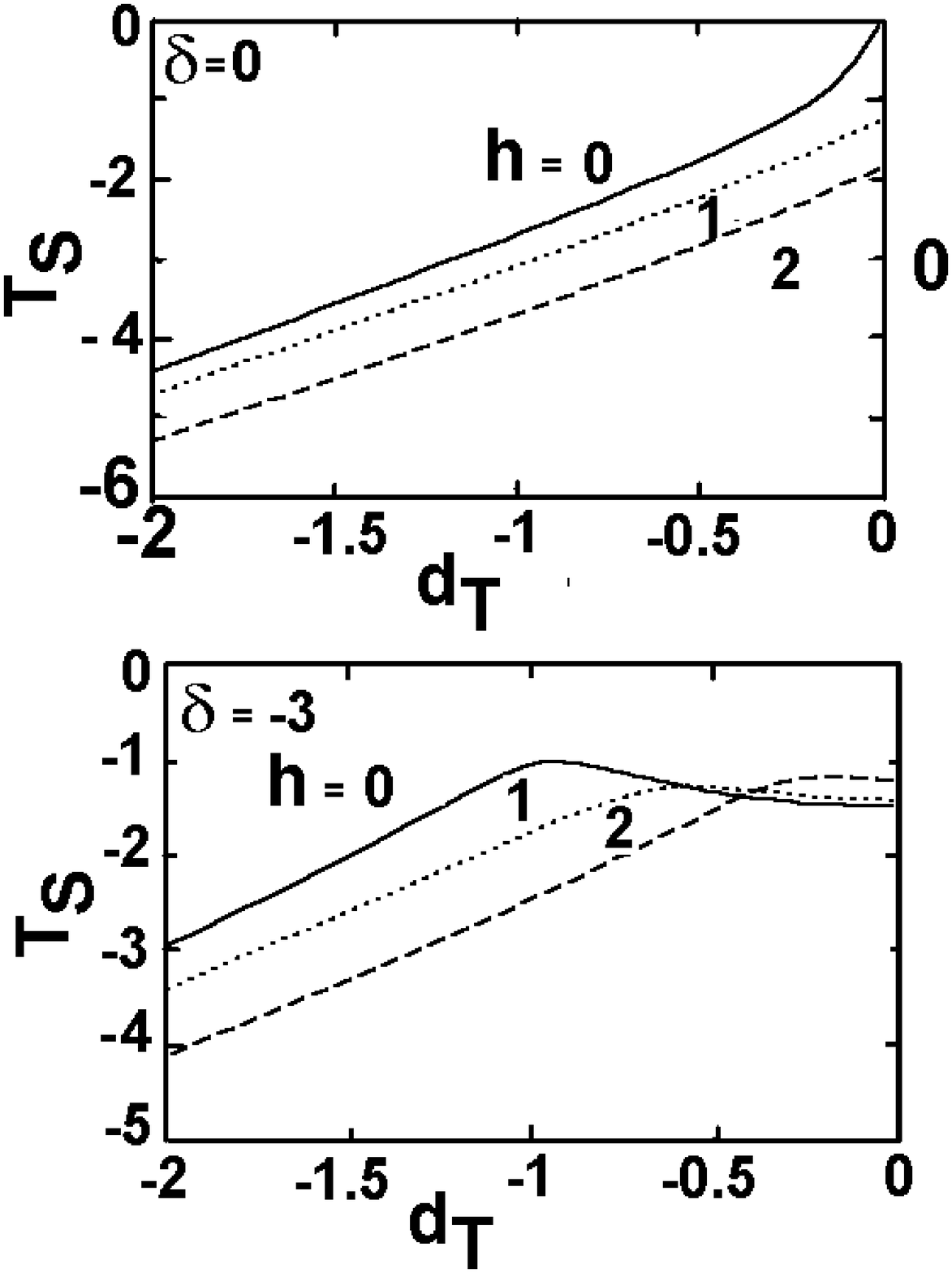}

\newpage

\epsfysize=4in
\epsffile{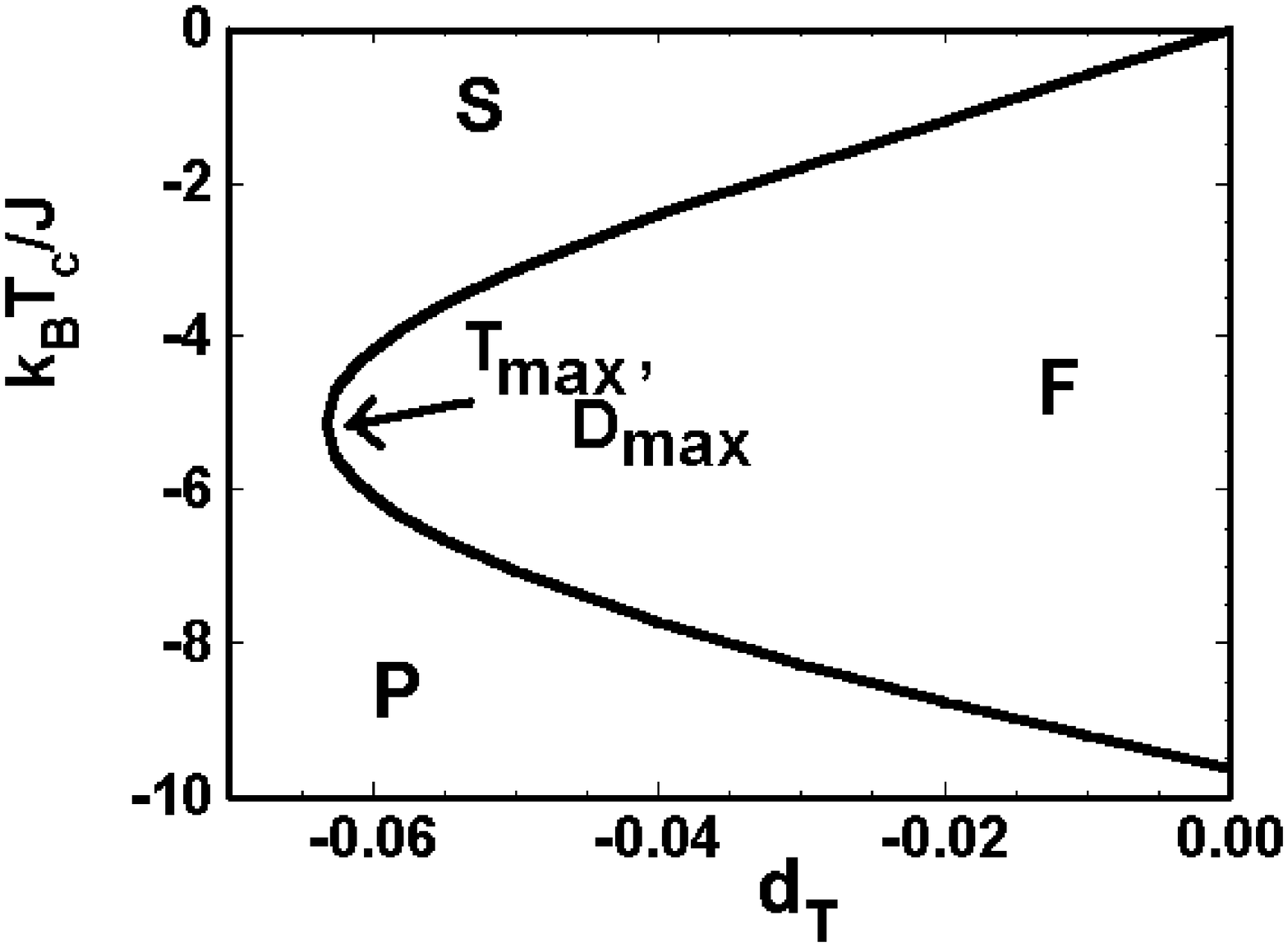}

\newpage

\epsfysize=5in
\epsffile{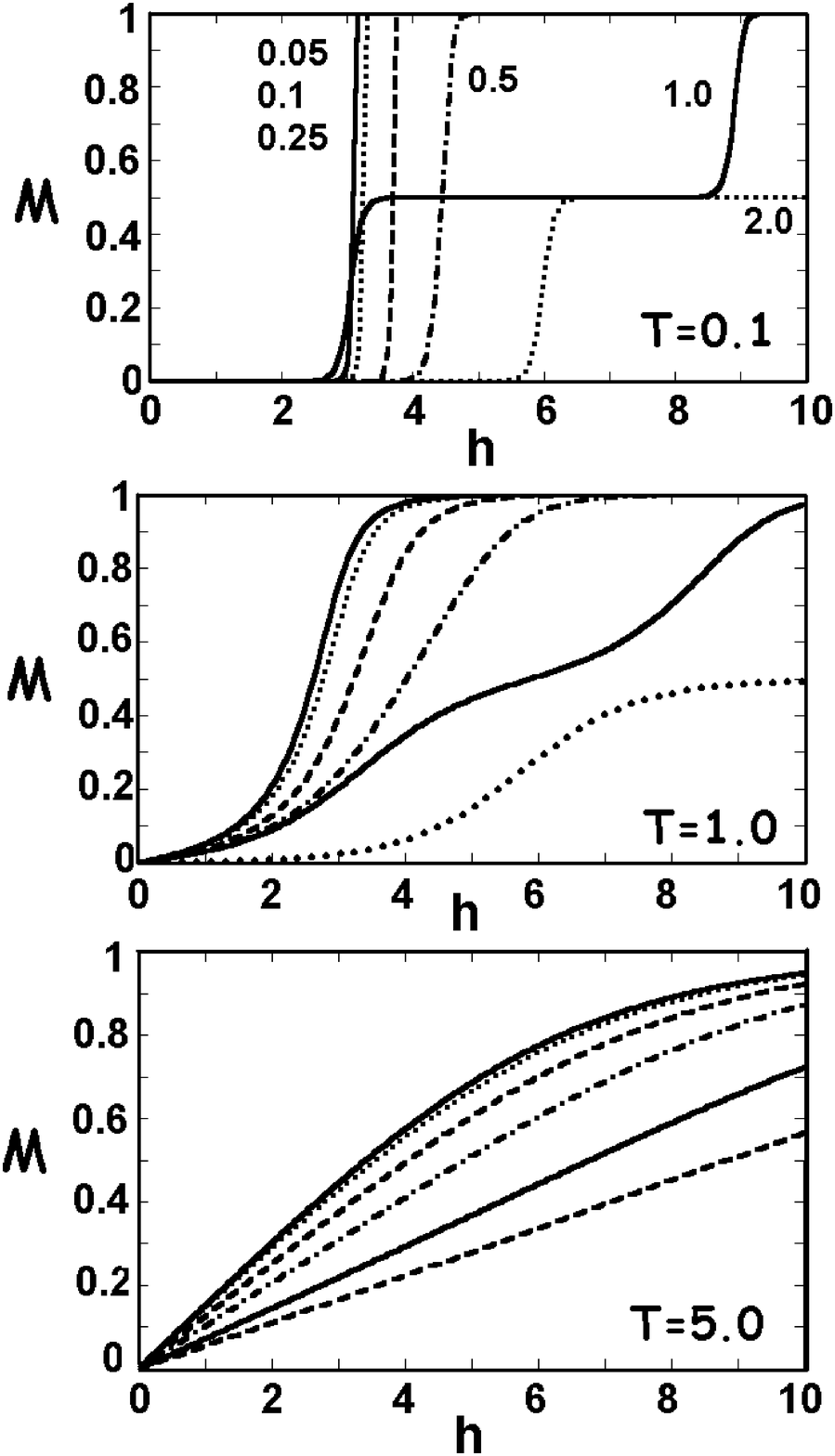}

\newpage

\epsfysize=5in
\epsffile{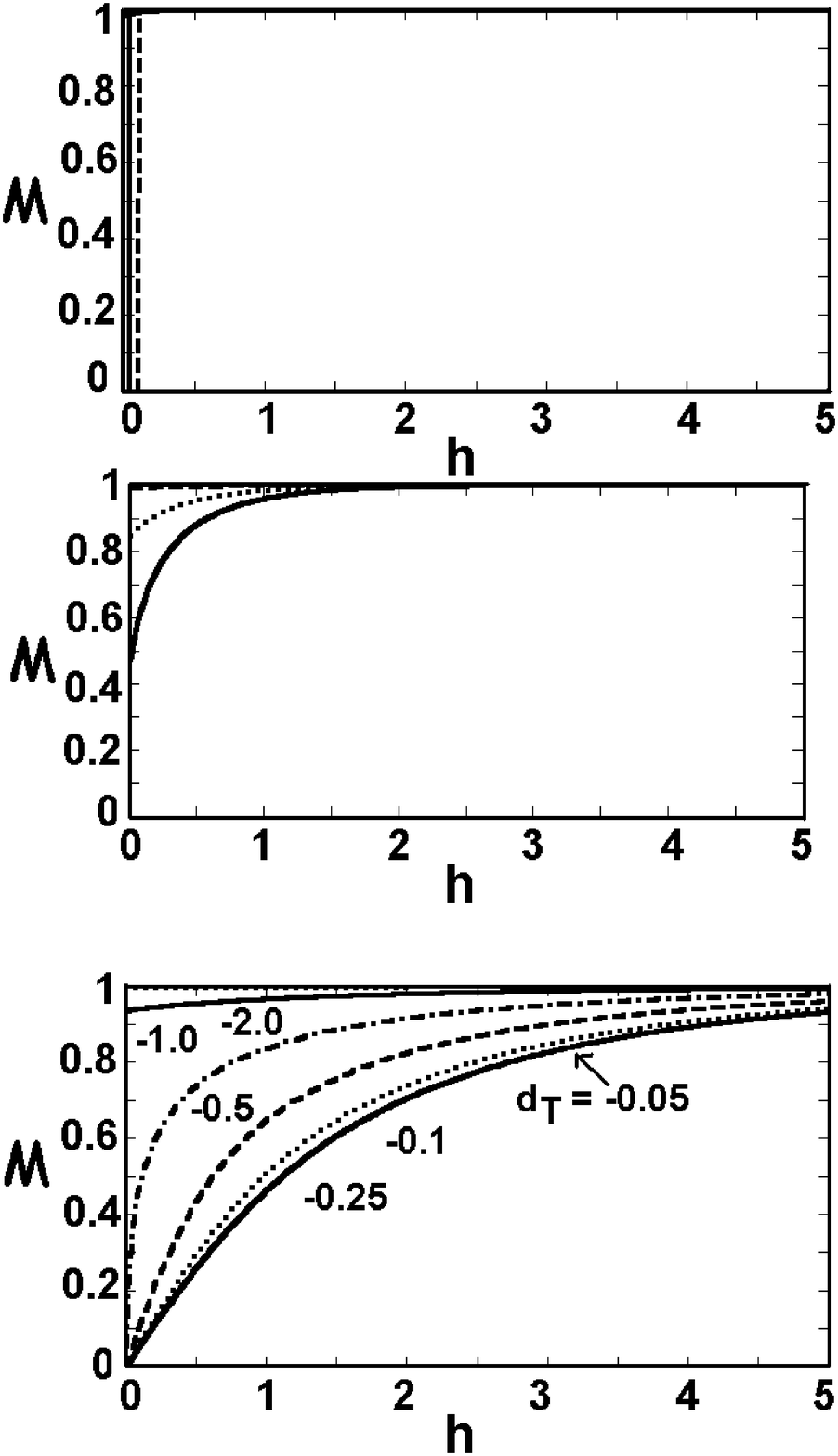}

\newpage

\epsfysize=5in
\epsffile{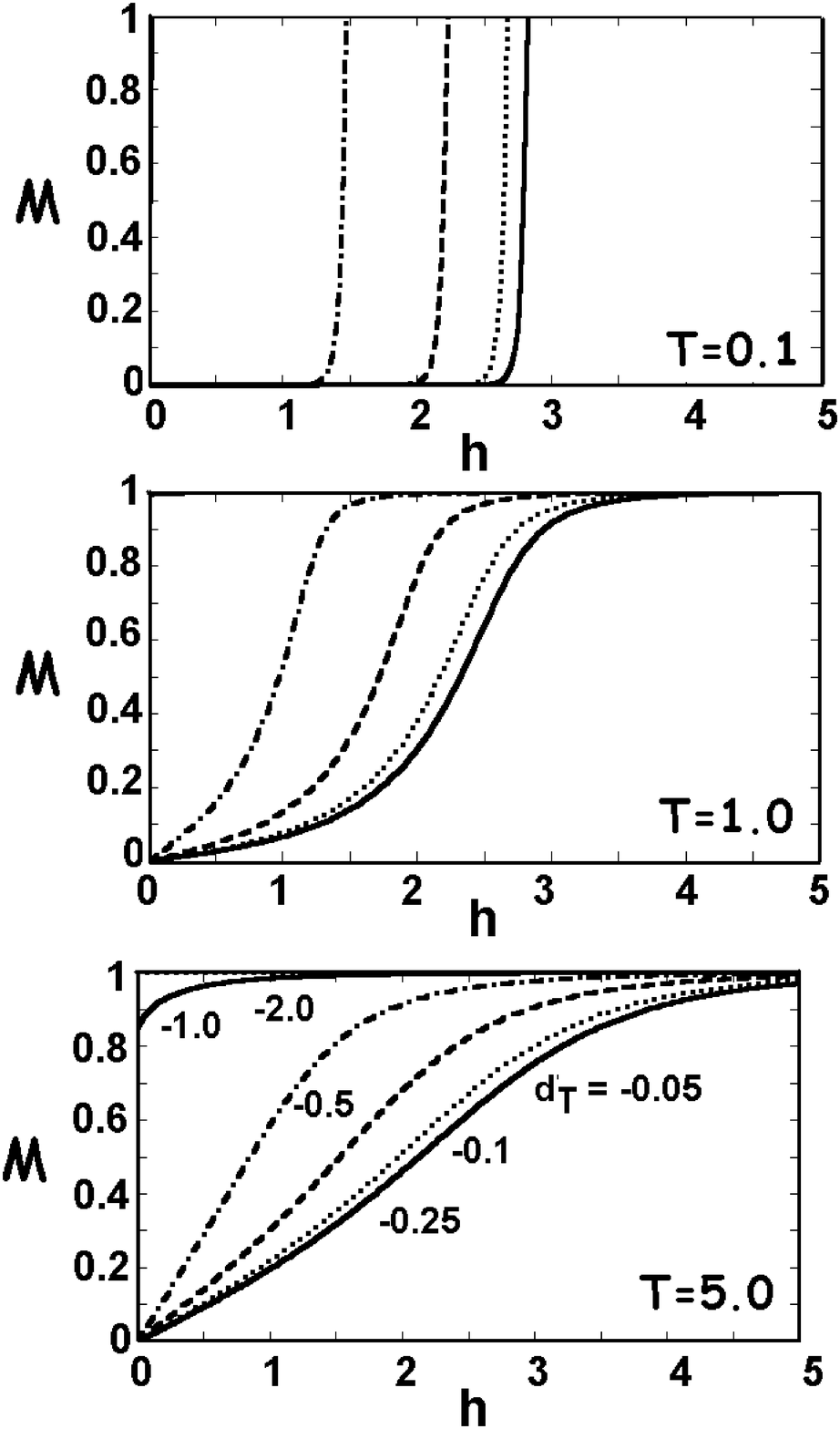}

\newpage

\epsfysize=5in
\epsffile{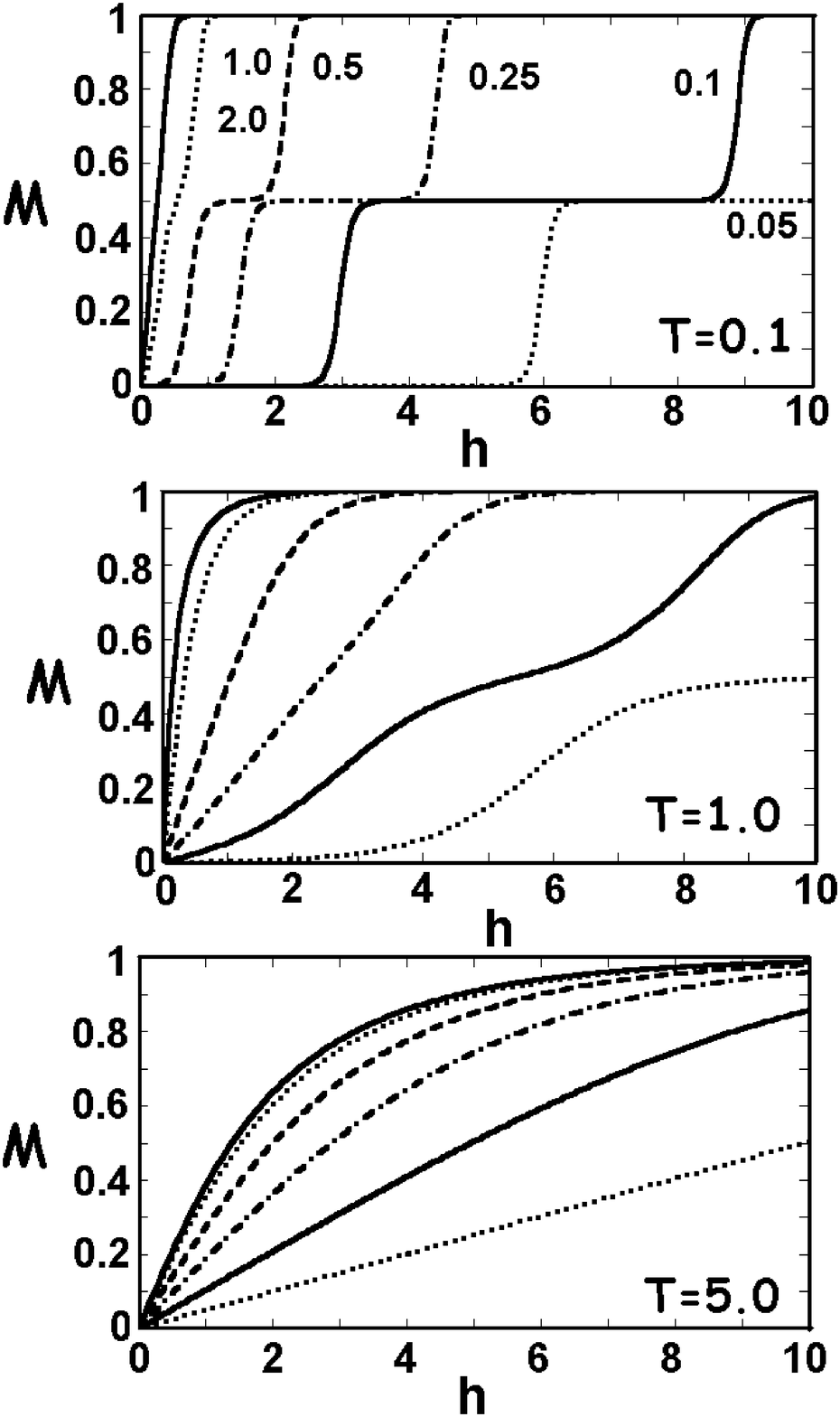}

\newpage

\epsfysize=5in
\epsffile{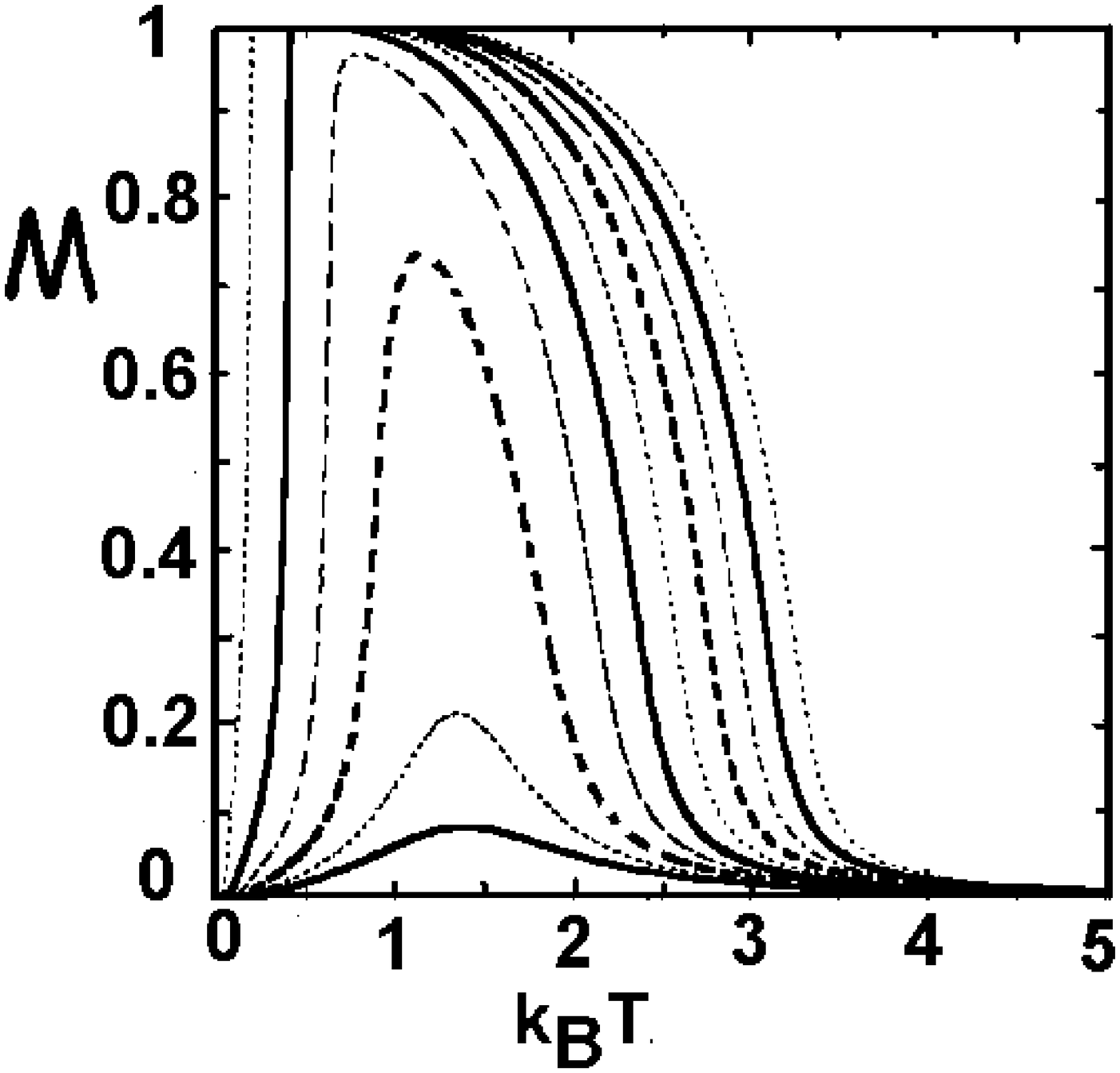}

\newpage

\epsfysize=4in
\epsffile{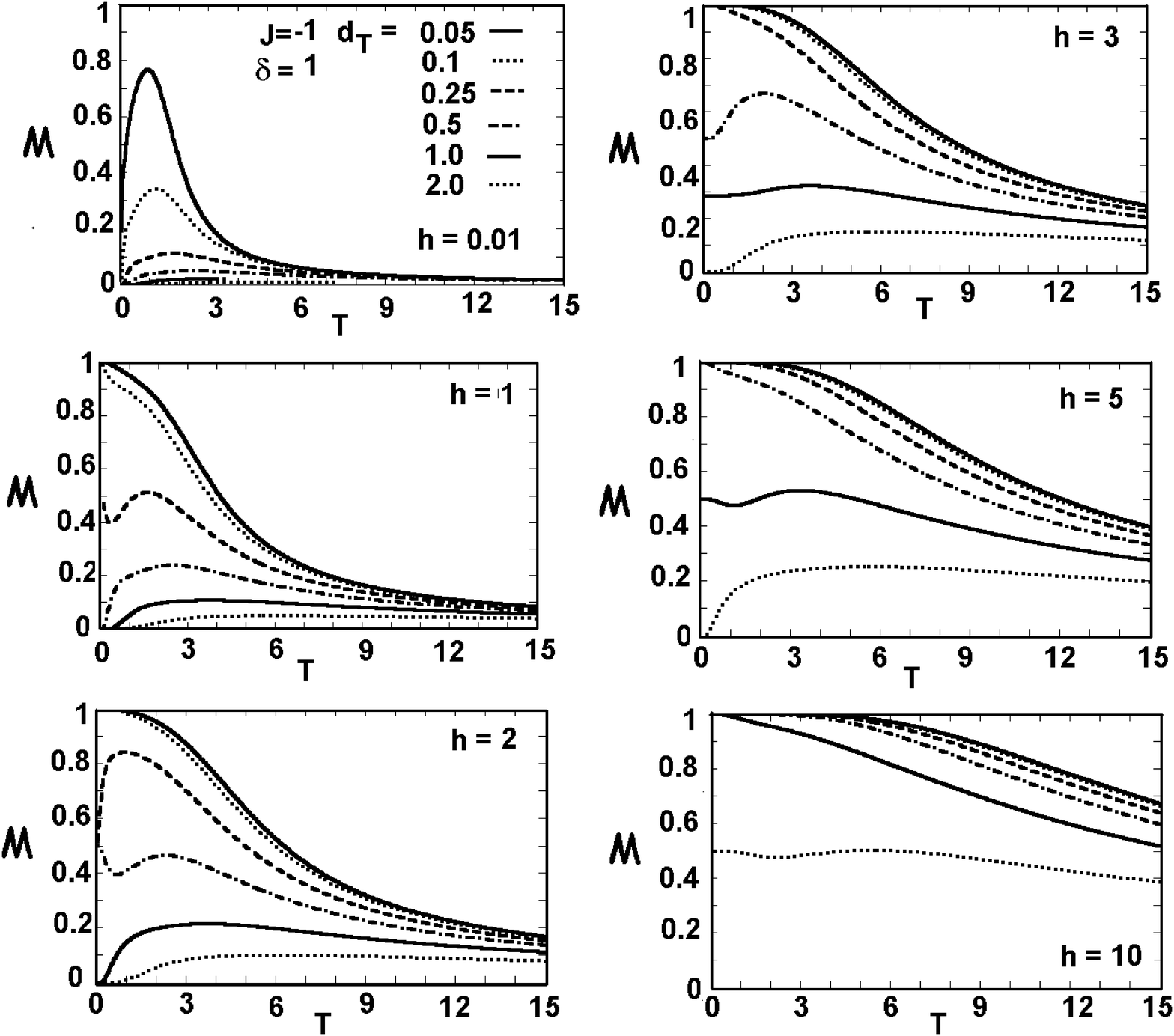}

\newpage

\epsfysize=5in
\epsffile{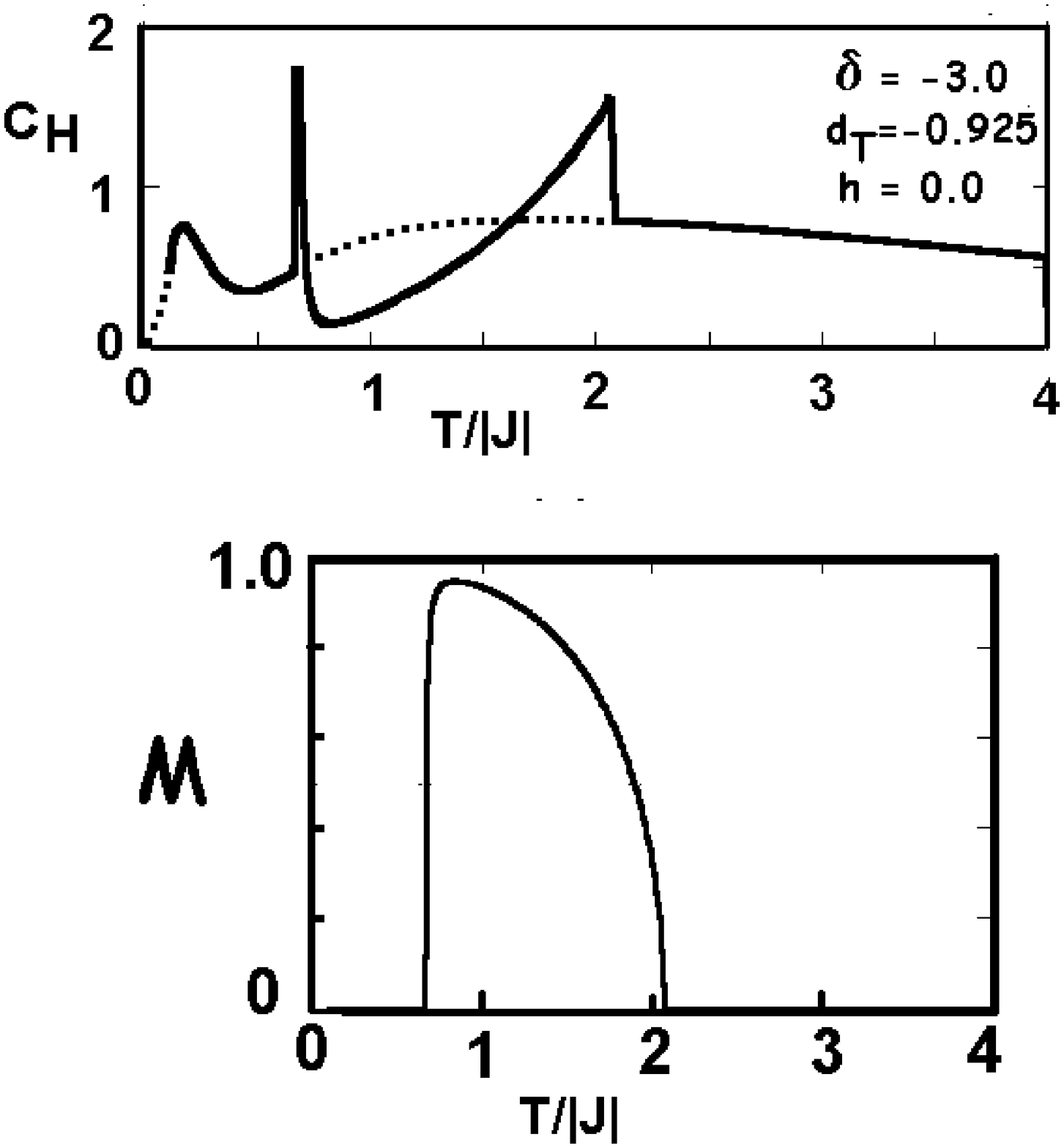}

\end{document}